\documentclass[longauth]{aa}  

\usepackage{graphicx}
\usepackage{txfonts}

\usepackage{natbib}
\bibpunct{(}{)}{;}{a}{}{,}            
\usepackage{amsmath}	
\usepackage{amsfonts}	
\usepackage{xcolor,colortbl}
\usepackage{color}

\usepackage[pdftex,unicode,psdextra,pdfencoding=auto,breaklinks=true,colorlinks=True]{hyperref}
\hypersetup{
    linkcolor=blue,  
    citecolor=blue,
    }

\usepackage{subfigure}

\newcommand{\tng}{{\sc{IllustrisTNG}}}
\newcommand{\eagle}{{\sc{eagle}}}
\newcommand{\skirt}{{\sc{skirt}}}

\begin{document} 

   \title{The S-PLUS Fornax Project (S+FP): Fornax-like clusters in cosmological hydrodynamical simulations}

\titlerunning{S+FP: Fornax-like clusters in cosmological simulations}

   \author{
        L.~J.~Zenocratti \inst{1,2}\fnmsep\thanks{Corresponding author: lzenocratti@fcaglp.unlp.edu.ar}
        \and
        A.~V.~Smith Castelli\inst{1,3}
        \and
        M.~E.~De Rossi\inst{4,5}
        \and
        D.~Pallero\inst{6,7}
        \and
        M.~C.~Artale\inst{8}
        \and
        R.~F.~Haack\inst{1,3}
        \and
        A.~R.~Lopes\inst{3,9}
        \and
        F.~R.~Faifer\inst{1,3}
        \and
        D.~Palma\inst{9}
        \and
        P.~K.~Humire\inst{9}
        \and
        J.~Thainá-Batista\inst{9}
        \and
        W.~Schoenell\inst{10}
        \and
        T.~Ribeiro\inst{11}
        \and
        A. Kanaan\inst{12}
        \and 
        C. Mendes de Oliveira\inst{9}
    }

   \institute{
        Facultad de Ciencias Astronómicas y Geofísicas, Universidad Nacional de La Plata, Paseo del Bosque s/n, La Plata, B1900FWA, Argentina
        \and 
        Departamento de Ciencias Básicas y Experimentales, Universidad Nacional del Noroeste de la provincia de Buenos Aires, Libertad 555, Junín, B6002GHF, Argentina
        \and
        Instituto de Astrofísica de La Plata (IALP), CONICET - UNLP, Paseo del Bosque s/n, La Plata, B1900FWA, Argentina
        \and
        Facultad de Ciencias Exactas y Naturales y Ciclo Básico Común, Universidad de Buenos Aires, Buenos Aires, Argentina
        \and
        Instituto de Astronomía y Física del Espacio (IAFE), CONICET – UBA, Buenos Aires, Argentina
        \and
        Departamento de Física, Universidad Técnica Federico Santa María, Avenida España 1680, Valparaíso, Chile
        \and
        Millennium Nucleus for Galaxies (MINGAL)
        \and
        Universidad Andres Bello, Facultad de Ciencias Exactas, Departamento de Fisica y Astronomia, Instituto de Astrofisica, Fernandez Concha 700, Las Condes, Santiago RM, Chile
        \and
        Departamento de Astronomia, Instituto de Astronomia, Geofísica e Ciências Atmosféricas da USP, Cidade Universitária, São Paulo, SP 05508-090, Brazil
        \and
        The Observatories of the Carnegie Institution for Science, 813 Santa Barbara St, Pasadena, CA 91101, USA
        \and
        Rubin Observatory Project Office, 950 N. Cherry Ave., Tucson, AZ 85719, USA
        \and
        Departamento de Física, CFM, Universidade Federal de Santa Catarina, PO Box 476, 88040-900, Florianópolis, SC, Brazil
    }

   \date{Received XXXX, 2026; accepted YYYY, 2026}

% \abstract{}{}{}{}{} 
% 5 {} token are mandatory
 
  \abstract
   {The Fornax cluster, the second-nearest rich galaxy cluster, constitutes a suitable laboratory to explore the evolution of galaxies in a dense environment. Recently, the Southern Photometric Local Universe Survey (S-PLUS) has obtained unprecedented photometric information of Fornax, revealing new features regarding its galaxy populations and surrounding regions. In this context, simulations are invaluable tools to interpret the past, present and fate of such observational findings.}
   {We aim to deliver a robust photometric catalog of simulated Fornax-like systems in cosmological context, to consistently contrast them with S-PLUS data. Such comparison will lead to our long-term goal: to trace the origin of Fornax galaxy populations and their environment.}
   {We analyze Fornax analogs from the \eagle\ and \tng\ simulations, selected using the observed properties of the Fornax cluster and its central galaxy NGC~1399. For each system, we generated synthetic photometry in the 12 S-PLUS bands using the \skirt\ radiative transfer code, reproducing the instrumental configuration of the S-PLUS survey. Simulated data cubes, mock images, spectral energy distributions, magnitudes and colors were obtained for each galaxy in our selected simulated Fornax analogs.
   }
   {The synthetic photometry and spectra derived from simulations show a good agreement with the S-PLUS observations, especially in the case of the central galaxy NGC~1399. We identify particular systems which show some similarity with the spatial distribution of galaxies in Fornax. Such simulated candidates reproduce the observed color-magnitude relation and the spatial substructure between the cluster core and the Fornax~A region. Also, simulated galaxies are bluer at higher cluster-centric distances, in agreement with observations. Although modest discrepancies were obtained between the observed and simulated color-magnitude diagrams in some cases, our results support the suitability of our selection criteria and synthetic photometry, and the reliability of current cosmological simulations to reproduce key general features of the Fornax cluster. 
   }
   {}


   \maketitle
\nolinenumbers 

%-------------------------------------------------------------------

\section{Introduction}

The Fornax cluster, the second-nearest rich galaxy cluster after Virgo, provides an excellent laboratory for studying galaxy formation and evolution in dense environments. It is centered on the elliptical galaxy NGC\,1399 and located at a distance of 20~Mpc \citep{2009Blakeslee} in the southern sky. Despite its relatively low dynamical mass ($M_{\rm dyn}\approx7 \times 10^{13}$ M$_\odot$; \citealp{2001Drinkwater}), Fornax stands out as a dynamically complex system, with a core dominated by early-type galaxies (ETGs), asymmetric features in its outskirts \citep{Iodice2019a}, and clear evidence of ongoing assembly through infalling groups such as the NGC\,1316 (a.k.a. Fornax A) group. Its mass, proximity, and rich multiwavelength coverage make it an ideal target for investigating the environmental processes that regulate galaxy formation. 

Observational efforts have provided a detailed view of Fornax and its galaxy population. Using optical imaging from the Fornax Deep Survey (FDS; \citealp{Peletier2020}), \citet{Iodice2019} revealed extended low-surface-brightness structures and intra-cluster light features around NGC\,1399, suggesting ongoing tidal interactions and accretion events. \citet{Venhola2022} presented a comprehensive catalog of dwarf galaxies from the same survey, finding a strong morphology-density relation and evidence of environmental quenching even in the cluster outskirts. ALMA observations by \citet{Zabel2019} detected molecular gas in several Fornax galaxies, indicating a diversity of gas depletion histories across different environments. In addition, \citet{Loni2021} combined VLT/MUSE and FDS data to study ionized gas in Fornax members, uncovering disturbed gas morphologies and signatures of ram-pressure stripping. Collectively, these works depict Fornax as a dynamically evolving system where tidal interactions and environmental effects continue to shape galaxy properties.

In this context, the Southern Photometric Local Universe Survey (S-PLUS; \citealp{Oliveira2019}) offers a unique opportunity to revisit Fornax from a complementary perspective. S-PLUS is an ongoing wide-field, 12-band imaging survey covering $9300~{\rm deg}^2$ of the southern sky, with most of its footprint already observed. As part of the main S-PLUS survey, images of the Fornax cluster covering $\sim 208$ square degrees around NGC\,1399 have been obtained. Within the S-PLUS Fornax Project (S+FP; \citealt{SmithCastelli2024}), these data are being used to carry out a comprehensive study of the Fornax cluster, analyzing its galaxy population and environmental structures across the entire surveyed area.
  
Interpreting such observational information requires the complementary use of cosmological simulations, which provide the physical context for understanding how clusters like Fornax assemble within the large-scale structure of the Universe. The interplay between modeling and observations has been central to extragalactic astronomy for decades. Numerical simulations have provided valuable insights into the formation and evolution of the Fornax cluster and its galaxy population. For instance, \citet{Bekki2003,Bekki2003b} modeled the dynamical evolution of nucleated dwarf galaxies, showing that tidal stripping within the Fornax potential can lead to the formation of ultra-compact dwarfs and extended globular cluster systems around massive galaxies. \citet{Harris2017} reproduced the diffuse intra-cluster light of Fornax, finding that it originates mainly from the disruption of dwarf galaxies. Hydrodynamical modeling of the infall of NGC\,1404 by \citet{Sheardown2018} successfully explained the observed X-ray morphology as the result of ram-pressure stripping during a high-velocity passage through the cluster core. Other studies have explored the origin of gas metallicity gradients \citep{Lara-Lopez2022}, the signatures of ancient massive mergers in Fornax galaxies \citep{Zhu2022}, the cluster luminosity function \citep{Venhola2022}, and the presence of dark matter in Fornax dwarf galaxies \citep{Asencio2022}. More recent works have identified Fornax-like systems within large-scale volumes, exploring the survival of stellar discs \citep{Galan2022,Ding2023}, the evolution and properties of HI gas \citep{Serra2023,Chaturvedi2024}, and the physical quenching mechanisms in dense environments \citep{Romero-Gomez2024}. Together, these studies depict Fornax as a dynamically young and still assembling system, offering an ideal benchmark for connecting simulations with multi-band observations such as those provided by S-PLUS.

In this framework, and within the S+FP, we aim to use cosmological simulations to identify and characterize Fornax-like systems, providing a theoretical counterpart to the S-PLUS observations. By combining observational and simulated data, we seek to uncover the physical drivers behind the observed trends in galaxy properties across the cluster environment. This paper is the first in a series dedicated to linking S-PLUS data with cosmological simulations. Here, we employ a set of publicly available simulations with different box sizes, resolutions, and subgrid physics implementations, allowing us to assess how these factors may influence our results. Specifically, we analyze clusters extracted from the \eagle\ and \tng\ simulations \citep{Schaye2015,Nelson15}. The main goal of this paper is to define and introduce a simulated sample of Fornax-like clusters and to describe the methodology used to generate synthetic observations of these systems for comparison with observations. A detailed physical interpretation and analysis, including the morphology, kinematics, star formation histories, and chemical enrichment of galaxies in Fornax and its simulated analogs, will be presented in future papers.

This work is organized as follows. In Section~\ref{sec:Fornax-SPLUS}, we present the observational data used to compare with the simulation results. In Section~\ref{sec:numerical_simulations}, we describe the simulations employed, the process used to identify galaxies and galaxy clusters within them, our selection criteria for identifying simulated Fornax-like clusters, and the methodology adopted to generate mock observations of galaxies within our Fornax-like candidates. In Section~\ref{sec:results}, we present the simulated observations and a brief comparison with S-PLUS observations of Fornax, focusing in particular on the best Fornax-like candidate in each simulation. In Section~\ref{sec:discussion}, we discuss our selection criteria, some caveats and differences between the simulated and observed data, and several potential studies that can be developed using our Fornax-like clusters and the simulated quantities we computed. Finally, in Section~\ref{sec:conclusions}, we summarize our work and outline future studies we plan to carry out.

%--------------------------------------------------------------------

\section{Observational data}
\label{sec:Fornax-SPLUS}

Fornax has been the focus of several complementary, multi-wavelength studies. In the optical range, it has been targeted by the ACS Fornax Cluster Survey (ACSFCS; \citealp{Jordan2007}), the Next Generation Fornax Survey (NGFS; \citealp{Munoz2015}), and the FDS (\citealp{Peletier2020}). There are also infrared observations from the Herschel Fornax Cluster Survey (\citealp{Davies2013}), radio studies such as the ALMA Fornax Cluster Survey mapping the cold molecular gas (\citealp{Zabel2019}), and the MeerKAT Fornax Survey, which provides HI observations (\citealp{Serra2023}). These efforts have focused on galaxies within the main cluster and the NGC\,1316 subgroup. However, a broader spatial coverage is required to assess the cluster surroundings and gain deeper insights into the scale and strength of environmental effects on galaxy evolution. 

The main S-PLUS survey in the Fornax area covers $\sim23 \times 11$ deg$^{2}$ across 106 fields, extending to $\sim5$ virial radii ($R_{\rm vir} \sim 700~\rm{kpc}$, \citealp{2001Drinkwater}) of Fornax along right ascension and up to $\sim3.5~R_{\rm vir}$ in declination. The observations were conducted with the T80-South, a 0.8-meter robotic telescope located at Cerro Tololo Inter-American Observatory (CTIO) in Chile. This telescope has a $1.4 \times 1.4$ deg$^2$ field of view and an imaging scale of 0.55 arcsec pixel$^{-1}$. The survey employs the Javalambre filter system \citep{2019/Cenarro}, which comprises five SDSS-like broad-band filters ($u$, $g$, $r$, $i$, $z$) and seven narrow-band filters ($J0378$, $J0395$, $J0410$, $J0430$, $J0515$, $J0660$, $J0861$), designed to sample key spectral features across the optical range. This filter configuration allows a simultaneous analysis of stellar and nebular emission components. For example, at the distance of Fornax (assuming a mean radial velocity of $\langle V_\mathrm{r} \rangle =1442$ km s$^{-1}$ around NGC\,1399; \citealp{Maddox2019}), the H$\alpha$+[NII] emission line falls within the $J0660$ filter (\citealp{2025Lopes}), while the five blue narrow bands provide valuable constraints on the young stellar population (\citealp{Thaina2023}).

Given the proximity of Fornax, the photometry must be optimized to detect and characterize a wide range of galaxy types, including large bright galaxies, dwarfs, compact objects, and low-surface-brightness systems. Therefore, within the context of the S+FP, we obtained a specific photometric catalog by running \texttt{SExtractor} in dual-mode on detection images created by combining the $g$, $r$, $i$, and $z$ bands of S-PLUS (see \citealt{Haack2024} for details). This catalog contains photometry in the 12 S-PLUS bands for $\sim 3 \times 10^6$ objects (both extended and compact) located in the direction of the Fornax cluster, covering an area of $\sim 208~{\rm deg}^2$ around NGC\,1399. From a compilation by \citet{SmithCastelli2024}, we also obtained a Fornax galaxy sample of 1,005 objects reported in the literature as members, either spectroscopically confirmed (23\%) or identified as likely members based on morphology. Crossmatching this sample with the S+FP catalog yields 464 galaxies with S-PLUS photometry, including 210 with available radial velocity measurements. In this work, we use this sample to compare our results from numerical simulations.

%--------------------------------------------------------------------

\section{Numerical simulations}
\label{sec:numerical_simulations}

\subsection{\eagle\ and \tng\ simulations}

The Evolution and Assembly of GaLaxies and their Environment (\eagle{\footnote{\url{https://icc.dur.ac.uk/Eagle/}}}; \citealp{Schaye2015,Crain2015}) and the Illustris The Next Generation (\tng{\footnote{\url{https://www.tng-project.org/}}}; \citealp{Marinacci18,Naiman18,Nelson2018,Pillepich18a,Springel18}) are suites of state-of-the-art cosmological hydrodynamical simulations in a flat $\Lambda$CDM cosmology. \eagle\ adopts cosmological parameters consistent with \citealp{Planck2014} ($\Omega_\Lambda = 0.693$, $\Omega_m = 0.307$, $\Omega_b = 0.048$, $h = 0.6777$, $\sigma_8 = 0.8288$, and $n_s = 0.9611$), while \tng\ assumes a cosmology consistent with \citealp{Planck16} ($\Omega_\Lambda = 0.6911$, $\Omega_m = 0.3089$, $\Omega_b = 0.0486$, $h = 0.6774$, $\sigma_8 = 0.8159$, and $n_s = 0.9667$). Both suites incorporate comprehensive and physically motivated galaxy formation models, implemented in a subgrid fashion, including detailed prescriptions for the main processes involved in galaxy formation and evolution (see \citealp{Schaye2015} and \citealp{Pillepich18a} for in-depth descriptions of the \eagle\ and \tng\ subgrid models, respectively). In addition, \tng\ incorporates ideal magnetohydrodynamics consistent with current cosmological estimates of the primordial magnetic field of the Universe (\citealp{Planck16}).

All simulations in \eagle\ and \tng\ begin at redshift $z=127$, with glass-like initial conditions and second-order perturbations. They follow the joint evolution of dark matter, stars, and gas within comoving cosmological volumes with periodic boundary conditions. The \eagle\ suite includes four main volumes with different particle numbers and mass resolutions (\citealp{Schaye2015}). In this work, we used the reference \eagle\ model with the largest box size and intermediate mass resolution, labeled RefL0100N1504, which has a box size of $(100~\rm{cMpc})^3$, an initial number of $(1504)^3$ particles for both gas and dark matter, and mass resolutions of $1.81\times10^6~\rm{M}_\odot$ and $9.7\times10^6~\rm{M}_\odot$ for baryonic and dark matter particles. The \tng\ suite includes three main volumes at three different mass resolution levels, all implementing the same reference subgrid physics model (\citealp{Pillepich18a}). Here, we used the three volumes at their highest resolution level, labeled TNG50-1, TNG100-1, and TNG300-1 (hereafter TNG50, TNG100, and TNG300, respectively), with comoving box sizes of $(51.7~ \mathrm{cMpc})^{3}$, $(106.5~ \mathrm{cMpc})^{3}$, and $(302.6~ \mathrm{cMpc})^{3}$. The corresponding baryonic mass resolutions are $8.5 \times 10^4~ \mathrm{M_\odot}$, $1.4 \times 10^6~ \mathrm{M_\odot}$, and $1.1 \times 10^7~ \mathrm{M_\odot}$, while the dark matter mass resolutions are $4.5 \times 10^5~ \mathrm{M_\odot}$, $7.5 \times 10^6~ \mathrm{M_\odot}$, and $5.9 \times 10^7~ \mathrm{M_\odot}$, respectively.

Both \eagle\ and \tng\ adopt subgrid models with parameters calibrated to reproduce the observed galaxy stellar mass function at $z \approx 0$, as well as the stellar mass-size, stellar mass-halo mass, and stellar mass-black hole mass relations, as described in \citet{Crain2015} and \citet{Pillepich18a}, respectively. Structure identification is carried out through a two-step procedure using the friends-of-friends (FoF) and \textsc{subfind} algorithms \citep{Springel01, Dolag09} to identify dark matter halos and subhalos (galaxies), respectively. Halos are identified by a unique integer number ({\it{GroupNumber}}, or `$gn$' for short) at a given snapshot, and subhalos are then classified hierarchically. They are distinguished into central and satellite systems, assigning an integer number (known as {\it{SubGroupNumber}}, or `$sgn$') to each subhalo belonging to a given halo at a given snapshot. Centrals correspond to the primary subhalo located at the minimum of the FoF halo gravitational potential, typically traced by the most bound dark matter particle, and are labeled with ${SubGroupNumber}=0$. More details on halo and subhalo identification in \eagle\ and \tng\ can be found in \citet{Schaye2015} and \citet{McAlpine16}, and in \citet{Nelson18}, respectively.

\subsection{Sample of simulated Fornax-like clusters}
\label{subsec:fornax_like_clusters}

In order to identify a suitable sample of Fornax-like candidates to be compared with S-PLUS observations, we extracted halos and their corresponding subhalos from the publicly available catalogs of the \eagle\ and \tng\ databases. Considering that estimates of the Fornax virial mass reported in the literature, obtained using different methods and associated uncertainties, span the range $3\times10^{13}~\rm{M}_{\odot}\lesssim M_{\rm vir,\ Fornax}\lesssim 9\times10^{13}~\rm{M}_{\odot}$ (e.g., \citealp{2001Drinkwater,Iodice2019a,Maddox2019}), and in order to avoid overly restrictive criteria, we selected, as an initial sample of simulated Fornax-like clusters, halos at redshift $z=0$ with a virial mass in the range $10^{13}~{\rm{M}}_\odot \leqslant M_{\rm 200} \leqslant 10^{14}~{\rm {M}}_\odot$. We use $M_{\rm vir}\equiv M_{200}$ in the simulations, where $M_{200}$ is the total mass enclosed within a sphere of radius $R_{200}$, defined as the radius containing a mean density 200 times the critical density of the Universe. This virial mass criterion is the only global condition used to constrain our initial sample of simulated clusters. Given the known virial mass-virial radius relation, our Fornax-like clusters are expected to have virial radii similar to, or at least of the same order as, that of the Fornax cluster (i.e., $R_{\rm vir,\ Fornax}\approx 0.7 ~\rm{Mpc}$; e.g., \citealp{Maddox2019}).

\begin{figure*}[ht!]
    \centering
    \includegraphics[width=\linewidth]{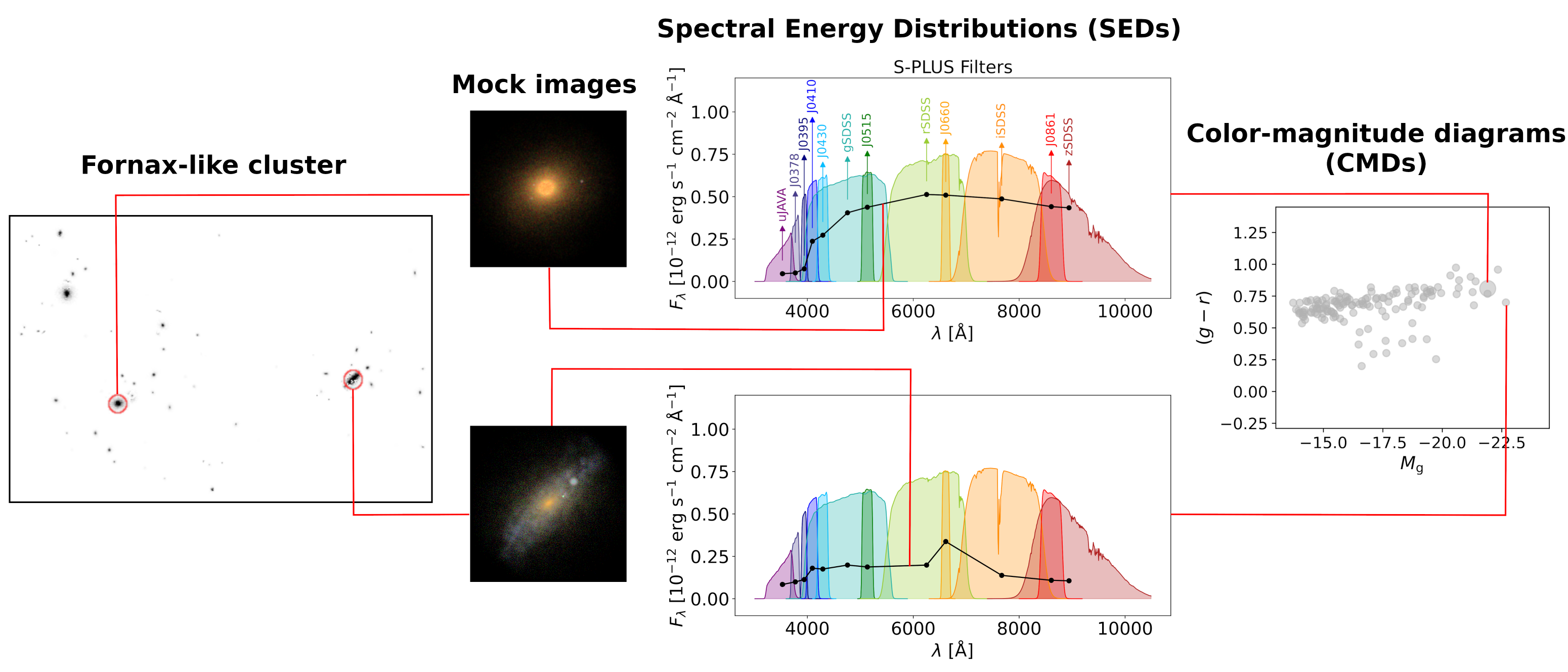}
    \caption{Simple schematic representation of our computation of simulated observations generated with the \skirt\ radiative transfer code.}
    \label{fig:scheme}
\end{figure*}

To refine our sample of Fornax-like clusters, and considering that selecting simulated clusters solely by their virial mass could lead to systems with properties different from those observed in Fornax, we further filtered the sample using the observed properties of NGC\,1399, the central galaxy of the cluster. In particular, we used the estimates from \citet{Iodice2019a} to constrain the stellar mass and radius of the central galaxy in our simulated clusters. In that work, the stellar masses $M_\star$ of ETGs in Fornax (including NGC\,1399) were estimated empirically by relating $\log(M_\star)$ to the $g-i$ color and the absolute $i$-band magnitude (\citealp{Taylor2011}, Eq.~8), with a $1\sigma$ accuracy of $\sim 0.10~\rm{dex}$. This yields a stellar mass for NGC\,1399 in the range $2.5\times10^{11}~\rm{M}_\odot \lesssim M_{\star,\ \rm{NGC\,1399}} \lesssim 4\times10^{11}~\rm{M}_\odot$. They also computed the effective radius of NGC\,1399 in the $u$, $g$, $r$, and $i$ bands, obtaining values of $23.5$, $35.1$, $31.2$, and $25.1~\rm{kpc}$, respectively, corresponding to an average effective radius of $\approx 30~\rm{kpc}$. Taking this into account, and considering that NGC\,1399 has an elliptical morphology, we selected from our initial sample those clusters whose central galaxy has a stellar mass in the range $1 \times 10^{11}~{\rm{M}}_\odot \leqslant M_\star \leqslant 5 \times 10^{11}~\rm{M}_\odot$, a stellar half-mass radius $R_{\rm{h},\star}$ within $20~{\rm{kpc}} \leqslant R_{{\rm h,}\star} \leqslant 40~{\rm{kpc}}$, and a visually spherical morphology, determined through visual inspection of edge-on and face-on renderings of the central galaxies extracted directly from the simulation databases. The adopted ranges in $M_\star$ and $R_{\rm{h},\star}$ were chosen to avoid overly restrictive selection criteria. With these criteria, we obtained 10 clusters from \eagle\ RefL0100N1504, 6 clusters from TNG100, and 29 clusters from TNG300.

With respect to the Fornax-like candidates in TNG50, no clusters fulfill all of our selection criteria. Nevertheless, since TNG50 is currently the highest-resolution simulation including galaxy clusters, we made an exception to select the systems most similar to Fornax from this simulation, following previous works \citep[e.g.,][]{Blana25, 2025Lopes}. Specifically, we did not apply the half-mass radius or visually spherical morphology criteria for the central galaxy to the TNG50 clusters. Without these restrictions, the simulation contains several clusters that are very similar to Fornax in terms of virial mass, virial radius, and number of members, and we therefore included them in our sample of simulated Fornax-like clusters. Therefore, our final sample also includes 14 clusters from TNG50. 

The complete list and main properties of our Fornax-like clusters are presented in Table~\ref{tab:clusters_dist_massive_galaxies}. They are primarily characterized by their {\it{GroupNumber}} ($gn$), virial mass ($M_{\rm 200}$), virial radius ($R_{\rm 200}$), number of member galaxies ($N$), and stellar mass of the central galaxy ($M_0$). Note that $N$ includes only galaxies with $M_\star \geqslant 5\times10^8~ \rm{M}_\odot$ (corresponding to systems with approximately 50, 500, and 5000 star particles in TNG300, \eagle\ RefL0100N1504 and TNG100, and TNG50, respectively), in order to avoid spurious results and numerical artifacts. Without this lower limit on $M_\star$, the number of subhalos identified by {\sc{subfind}} increases considerably relative to the number of galaxies in our Fornax-like clusters, but many of these subhalos should not be considered galaxies, as they may correspond to structures with little or no stellar content.

\subsection{The \skirt\ radiative transfer code}

For a consistent comparison between our Fornax-like clusters and S-PLUS observations of Fornax, we post-processed the simulations to compute synthetic photometric quantities compatible with S-PLUS data. We used the \skirt{\footnote{\url{https://skirt.ugent.be/root/_home.html}}} radiative transfer code (\citealp{Camps2020}) to generate synthetic spectra, magnitudes, and images (data cubes). This procedure was applied to every galaxy in our Fornax-like clusters in order to produce simulated observations that can be directly compared with S-PLUS results.

Our implementation of \skirt\ was mainly based on those of \citet{Trayford2017} and \citet{RodriguezGomez2019} for \eagle\ and \tng, respectively. We refer the interested reader to these works for complete technical details. However, unlike the implementations adopted in those studies, and in order to reduce computational cost and processing time, we ran \skirt\ assuming no dust in the simulated systems, thus considering only stellar radiation. The analysis of dust effects in simulated galaxies is beyond the scope of this work. Therefore, the simulated observational quantities presented here can be interpreted as observations already corrected for dust effects (e.g., absolute magnitudes and observed colors corrected for internal and external extinction and reddening). These simulated quantities were compared with observations already corrected for such effects (see Secs.~\ref{subsec:fornax_likes_seds} and \ref{subsec:fornax_likes_magnitudes}).

Since the simulated systems are intended to be analogs of the Fornax cluster, \skirt\ was configured so that the distance between the simulated instruments and the observed systems is consistent with the estimated distance to Fornax. A distance modulus of $(m-M)=31.51$ was adopted (\citealp{Iodice2019a}), representing the only observational constraint from Fornax used to generate the mock observations. From this distance modulus, the corresponding redshift, $z$, and luminosity distance to the simulated systems, $d_{\rm L}$, were estimated with \skirt, and these quantities were then used to compute the synthetic spectra and photometry.

The main goal of generating synthetic observations is to directly compare simulated and observed quantities. In this context, \skirt\ was configured to match the characteristics of the S-PLUS survey instruments (see Sect.~\ref{sec:Fornax-SPLUS} and \citealp{Oliveira2019} for details of the survey, and \citealp{SmithCastelli2024} for details of the S+FP project within which this comparison is framed). Specifically, we configured \skirt\ to simulate observations using the 12-filter S-PLUS system, adopting the corresponding transmission curve for each filter. As a result, synthetic images, spectra, and photometric magnitudes were generated in the 12 S-PLUS bands for every galaxy in our Fornax-like clusters. These quantities can therefore be directly compared with those obtained from S-PLUS observations. At each wavelength included in the adopted filter system, \skirt\ launched $4\times 10^6$ photon packets, a number that provides a reasonable balance between simulation accuracy and computational cost (\citealp{Trayford2017}). Fig.~\ref{fig:scheme} shows a schematic representation of the simulated observations generated with \skirt.

\begin{table*}[!ht]
\centering
\caption{Best Fornax-like clusters defined in this work.}
\small
{
\setlength{\tabcolsep}{8.8pt}
\renewcommand{\arraystretch}{1.} 
\begin{tabular}{ccccccccccc}
\hline
\hline
Simulation & $gn$ & $\log(M_{\rm 200})$ & $R_{\rm 200}$ & $\log(M_{\star,\rm 0})$ & $\log(M_{\star,\rm sec})$ & $d$ & $d_{\rm xy}$ & $d_{\rm xz}$ & $d_{\rm yz}$ & $N$ \\
 & & $[\rm{M_\odot}]$ & $[\rm{Mpc}]$ & $[\rm{M_\odot}]$ & $[\rm{M_\odot}]$ & $[\rm{Mpc}]$ & $[\rm{Mpc}]$ & $[\rm{Mpc}]$ & $[\rm{Mpc}]$ & \\
\hline
\eagle\ RefL0100N1504 & 24 & 13.68 & 0.77 & 11.67 & 10.54 & 0.62 & 0.62 & 0.11 & 0.62 & 50 \\
TNG100 & 71 & 13.39 & 0.61 & 11.58 & 10.72 & 0.78 & 0.74 & 0.57 & 0.59 & 53 \\
TNG300 & 246 & 13.76 & 0.81 & 11.66 & 11.66 & 2.26 & 2.25 & 1.27 & 1.88 & 138 \\
TNG50 & 2 & 13.81 & 0.85 & 12.22 & 11.02 & 1.03 & 1.00 & 0.25 & 1.02 & 141 \\
\hline
\end{tabular}
\tablefoot{From left to right, the columns list the simulation to which each cluster belongs, the cluster identifier within the corresponding simulation ($gn$), its virial mass ($M_{\rm 200}$) and virial radius ($R_{\rm 200}$), the stellar masses of the central and second most massive galaxies ($M_{\star,\rm 0}$ and $M_{\star,\rm sec}$, respectively), the 3D distance between these galaxies ($d$), their projected distances in the three coordinate planes ($d_{\rm xy}$, $d_{\rm xz}$, and $d_{\rm yz}$), and the number of member galaxies in the cluster ($N$). These clusters, as well as all our Fornax-like systems, were selected from the simulations at redshift $z=0$.
}
}
\label{tab:best_Fornax-likes}
\end{table*}

\begin{figure*}
\centering
\subfigure{\includegraphics[width=0.24\textwidth]{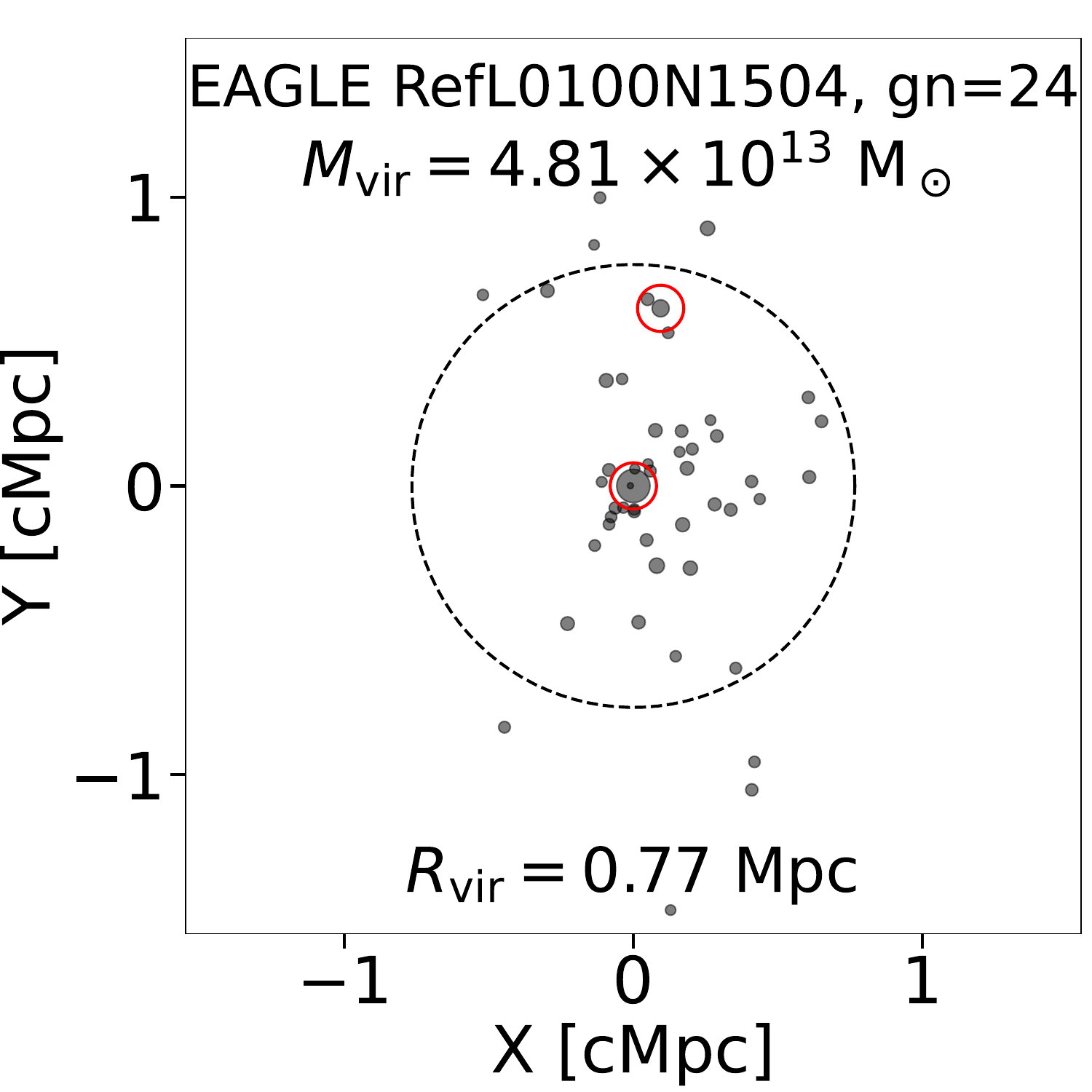}}
%\hspace{0.05\textwidth}
\subfigure{\includegraphics[width=0.24\textwidth]{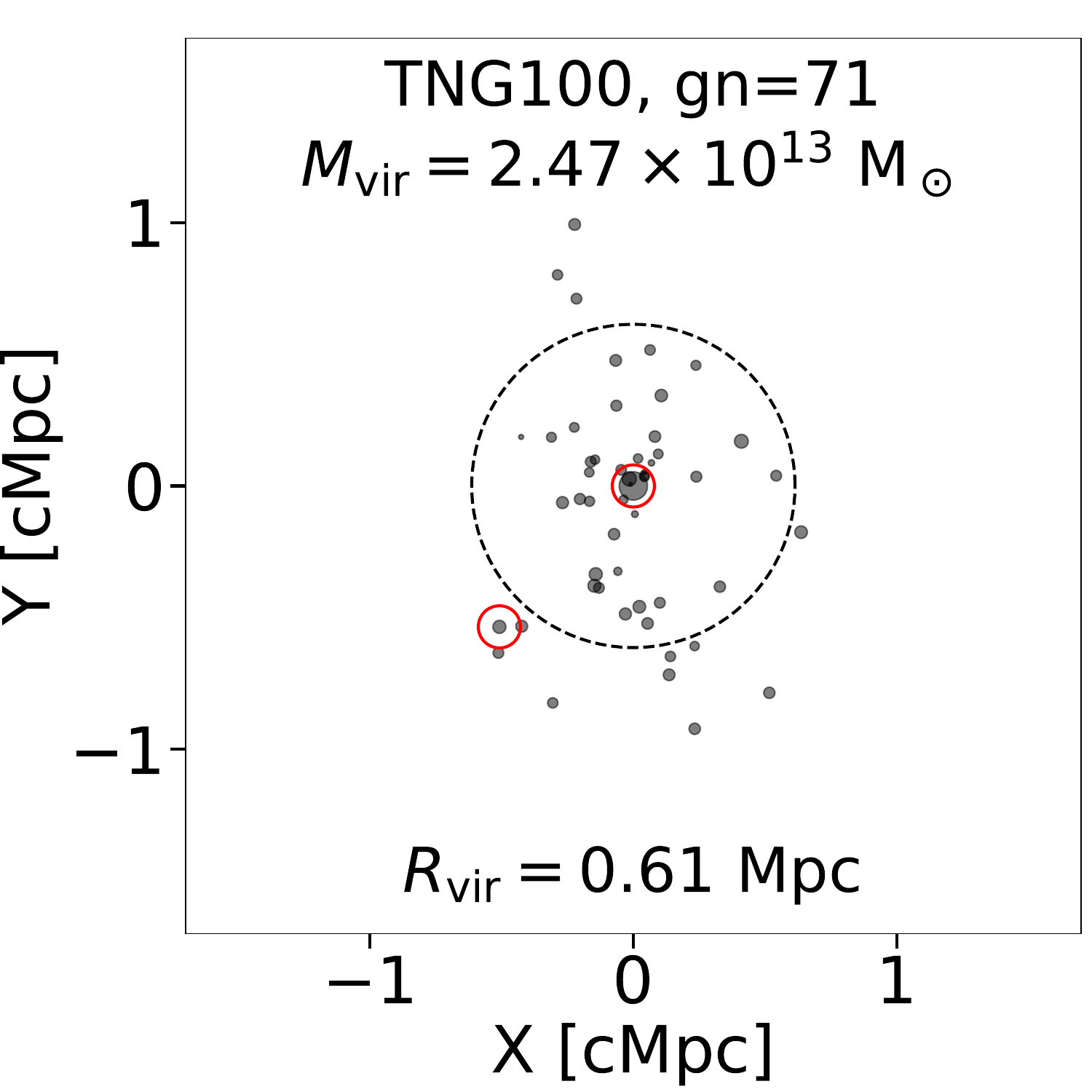}}
%\par
\subfigure{\includegraphics[width=0.24\textwidth]{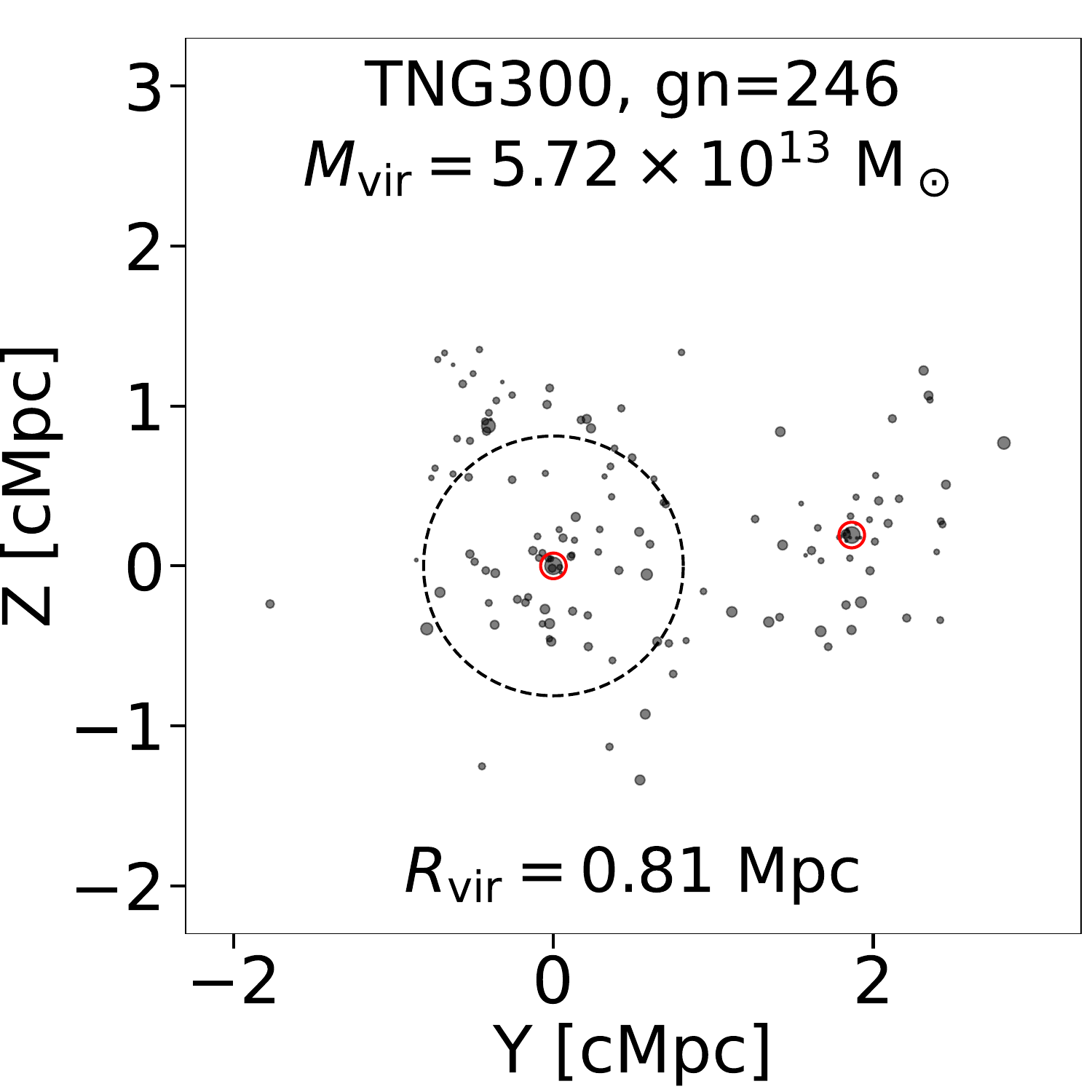}}
%\hspace{0.05\textwidth}
\subfigure{\includegraphics[width=0.24\textwidth]{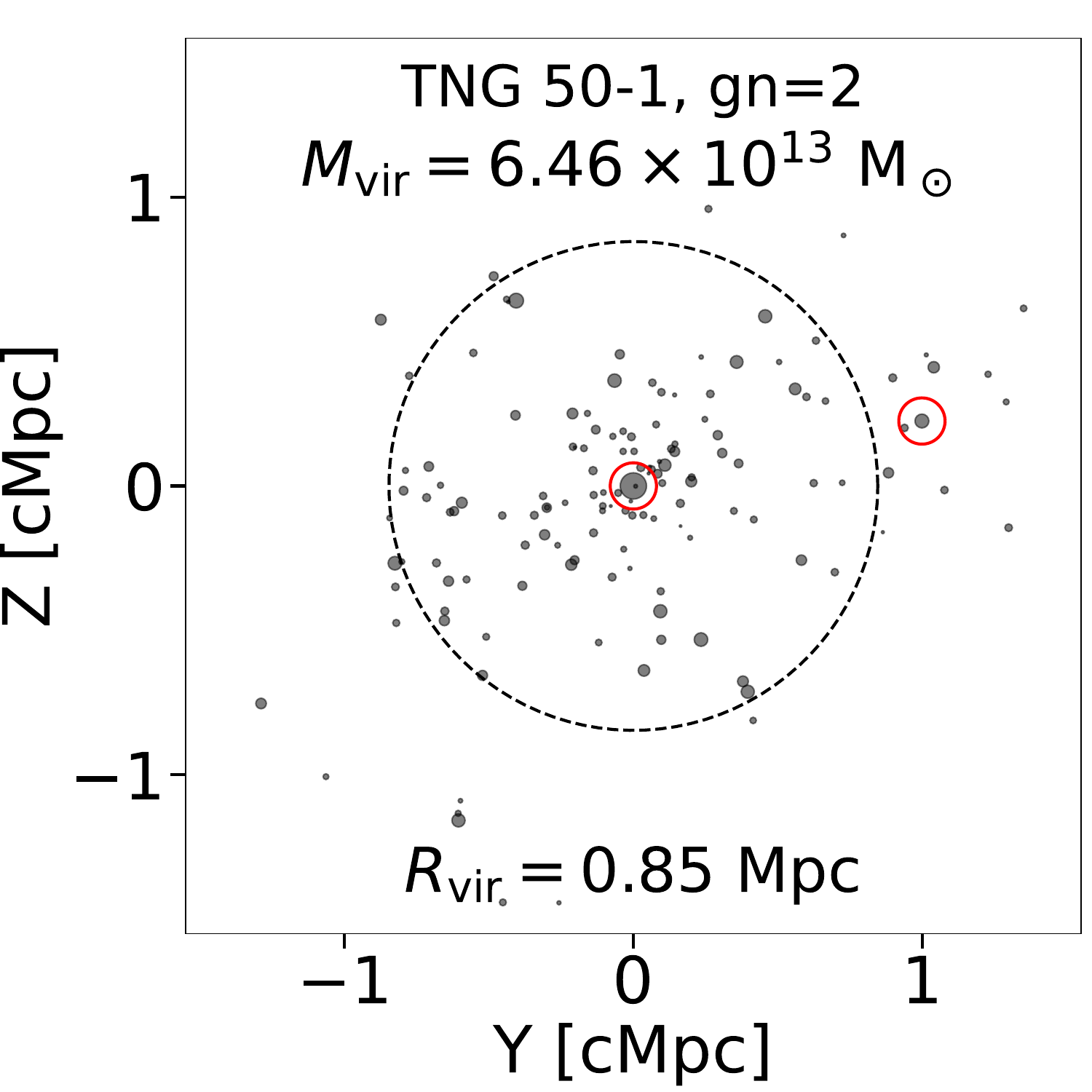}}
\caption{
Galaxy distribution in our best Fornax-like clusters. From left to right: the best Fornax-like candidates extracted from the \eagle\ RefL0100N1504, TNG100, TNG300, and TNG50 simulations, respectively. Each panel lists the corresponding simulation, the identifier of the group within the simulation ($gn$), its virial mass ($M_{\rm vir}$), and its virial radius ($R_{\rm vir}$). Black symbols represent member galaxies of each cluster. The black dashed circle encloses the corresponding virial radius. Red solid circles indicate the two most massive galaxies in the corresponding cluster (the central galaxy, located at the center of the virial radius circle, and the second most massive galaxy). Coordinates are given with respect to the corresponding cluster center.}
\label{fig:Fornax-likes_spatial_dist}
\end{figure*}

Finally, regarding the emission sources, as mentioned above, we only considered stellar sources extracted directly from the simulation databases. Note that ``stellar sources'' refers to star particles, which do not represent individual stars, but rather simple stellar populations (SSPs). For each simulated galaxy extracted from \eagle\ and \tng, we selected star particles within a $50~\rm{kpc}$ sphere centered on the potential minimum. Following the methodology of \citet{Trayford2017} and \citet{RodriguezGomez2019}, but in a simplified way, we divided the stellar sources according to their age, $t$, into very young stellar populations ($t\leqslant 10~\rm{Myr}$) and older stellar populations ($t>10~\rm{Myr}$). The very young populations were assigned spectral models from the {\sc{mappings-iii}} library (\citealp{Groves2008}), while the older populations were modeled using spectral energy distributions from the {\sc{galaxev}} models (\citealp{Bruzual2003}) with a \citet{Chabrier2003} initial mass function. These models are already implemented in \skirt\ and were internally configured for use by the code. The {\sc{mappings-iii}} models were chosen because they include emission lines and gas and dust absorption within star-forming clouds, allowing an accurate representation of active and recent star-forming regions. The choice of the {\sc{galaxev}} models for the older stellar populations is motivated by their straightforward implementation, the small number of stellar properties they require, and their widespread use in reproducing the optical properties of local galaxy populations with realistic star formation and chemical enrichment histories. The data required by \skirt\ to simulate the stellar sources and their modeling can be extracted directly from the simulation databases or calculated easily based on them (see Appendix~\ref{appendix:skirt}, and \citealp{Trayford2017} and \citealp{RodriguezGomez2019} for details).

%--------------------------------------------------------------------

\section{Results}
\label{sec:results}

\subsection{Galaxy distribution in simulated Fornax-like clusters}
\label{subsec:Fornax-like_spatial_distribution}

Our selection criteria allowed us to obtain simulated clusters with relevant properties similar to those of Fornax. In this section, we explore the predictions of the simulations regarding their galaxy spatial distribution, an aspect that was not considered when defining our sample. In particular, the Fornax cluster presents two main substructures: the central region, dominated in mass and brightness by NGC\,1399, and the NGC\,1316 subgroup region, with the latter being the second most massive galaxy in Fornax. To perform a robust comparison between Fornax observations and simulated clusters, it is important to note that the projected distance between NGC\,1316 and NGC\,1399 (i.e., between the NGC\,1316 subgroup and the center of the Fornax cluster) has been estimated to be $\sim 1.5~\rm{Mpc}$ (\citealp{Iodice2019}). Thus, the region dominated by NGC\,1316 lies outside the estimated virial radius of Fornax. Therefore, a reasonable simulated analog of Fornax should have its second most massive galaxy located outside the corresponding virial radius, at a distance of $\sim 1~\rm{Mpc}$ from the central galaxy. Naturally, not all of our simulated Fornax-like clusters are expected to satisfy this spatial distribution criterion. Those that do, even without a strict constraint on the distance between the two dominant components, may be considered the strongest analogs of the Fornax cluster.

As mentioned above, Table~\ref{tab:clusters_dist_massive_galaxies} lists all of our Fornax-like clusters, including the distance between their two most massive galaxies. Based on this information, we selected four simulated clusters (one from each simulation used) as our best simulated Fornax-like systems. These clusters, chosen as representative examples to illustrate the types of analyses and simulated observations that can be performed, are listed in Table~\ref{tab:best_Fornax-likes}, where each cluster is identified within its corresponding simulation by its $gn$. The table also includes the virial mass and virial radius of each cluster ($M_{\rm 200}$ and $R_{\rm 200}$, respectively), the stellar masses of the two most massive galaxies ($M_{\star,\rm 0}$ and $M_{\star,\rm sec}$, corresponding to the central galaxy and the second most massive galaxy of a given group), the 3D distance between these galaxies ($d$), their projected distances onto the three coordinate planes of the simulated cosmic box, and the number of galaxies $N$ within each simulated cluster.

In Fig.~\ref{fig:Fornax-likes_spatial_dist}, we show the spatial distribution of galaxies in our best Fornax-like clusters. The adopted projections were chosen to best highlight the similarity between the simulated clusters and the observed structure of the Fornax cluster, considering that any projection onto the coordinate planes can be interpreted as a random view in the simulations. In particular, the similarity between Fornax and the simulated cluster extracted from TNG300 is evident: the two most massive galaxies are separated by a considerable distance, and two main structures can be identified within the cluster, each centered around one of these dominant galaxies. The central region of this cluster is comparable to the central region of Fornax, dominated by NGC\,1399, while the region associated with the second most massive galaxy, located outside the cluster virial radius, resembles the NGC\,1316 subgroup region. Also noteworthy is the considerably higher number of member galaxies in this cluster compared with the rest of our best Fornax-like systems.

\subsection{Simulated data cubes and mock images}
\label{subsec:fornax_likes_mock_images}

We used the \skirt\ code to generate mock images of the galaxies in our Fornax-like clusters. These images are composed of 12 frames (one for each S-PLUS filter), adopting the same image scale as the S-PLUS camera, namely $0.55~\rm{arcsec},\rm{px}^{-1}$. Considering the distance to the Fornax cluster, a field of view of $17.36\times17.36~\rm{arcmin}^2$ was adopted to obtain this instrumental scale, corresponding to a projected square area of $(100~\rm{kpc})^2$ at the distance of Fornax, large enough to fully enclose each galaxy in the simulated Fornax-like clusters. The positional parameters (i.e., azimuthal, inclination, and position angles) were set to zero (their default values) in order to preserve the random orientations of the simulated galaxies. With the adopted image scale and field of view, each frame of the simulated data cubes corresponds to an image of $1870~\rm{px} \times 1870~\rm{px}$, where each pixel contains the corresponding observed surface brightness.

\begin{figure*}
\centering
\subfigure{\includegraphics[width=0.23\textwidth]{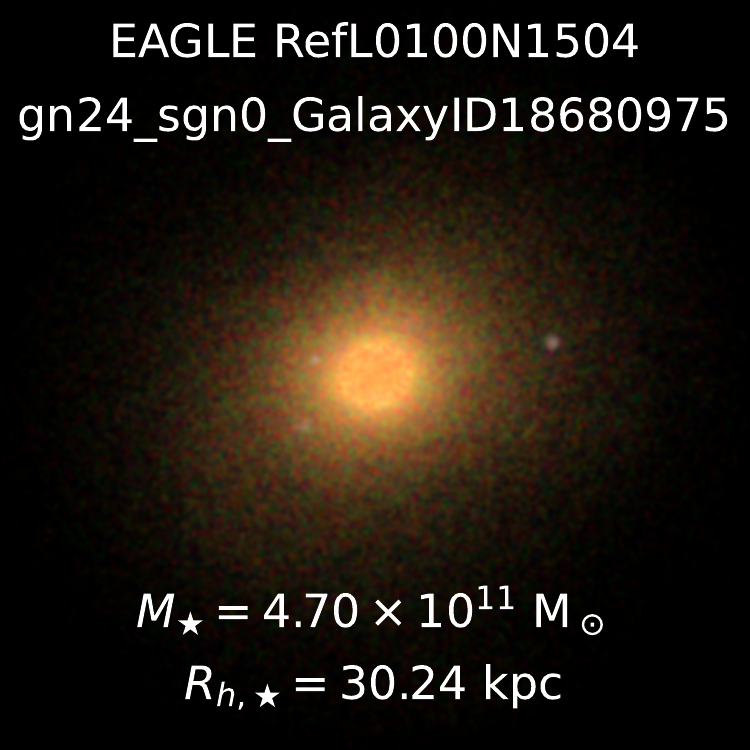}}
\subfigure{\includegraphics[width=0.23\textwidth]{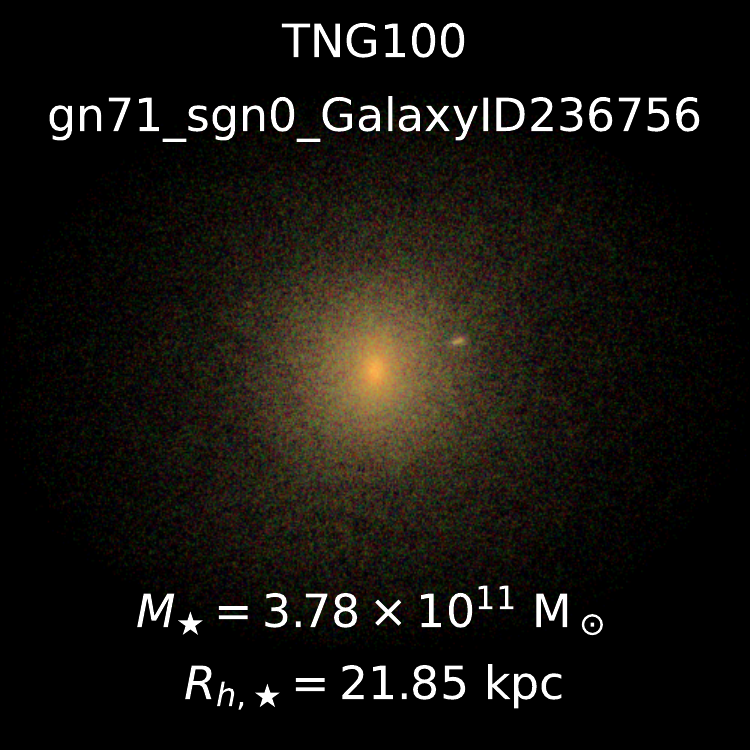}}
\subfigure{\includegraphics[width=0.23\textwidth]{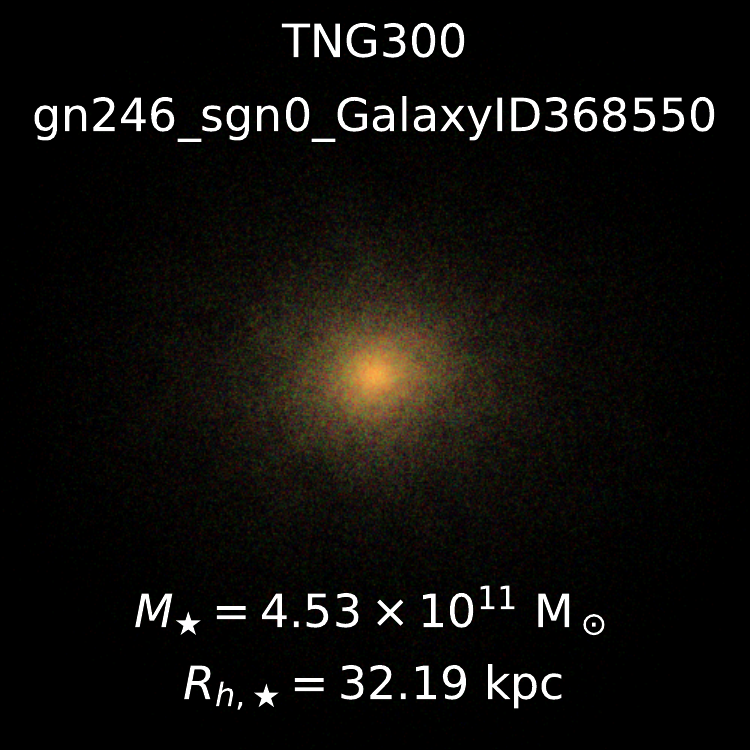}}
\subfigure{\includegraphics[width=0.23\textwidth]{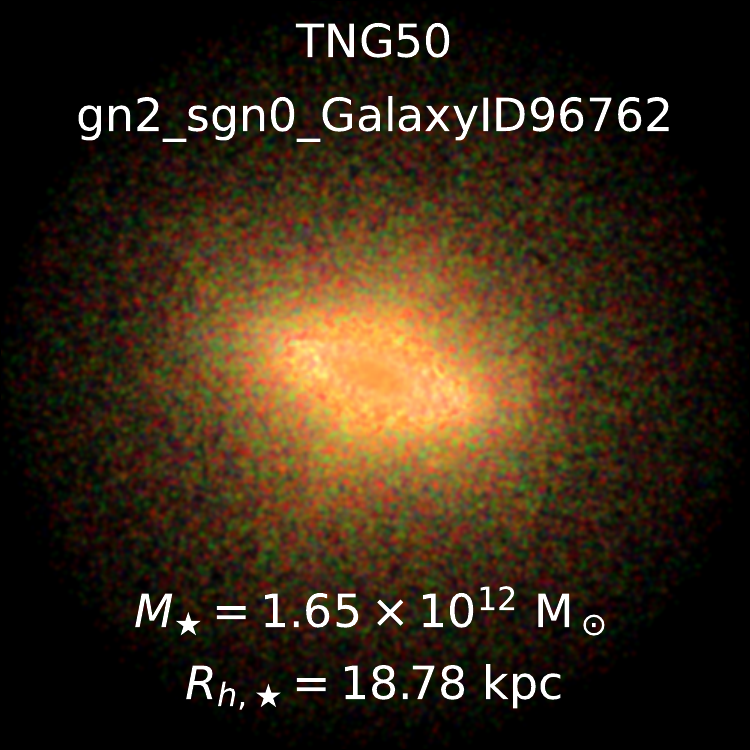}}
\par
\subfigure{\includegraphics[width=0.23\textwidth]{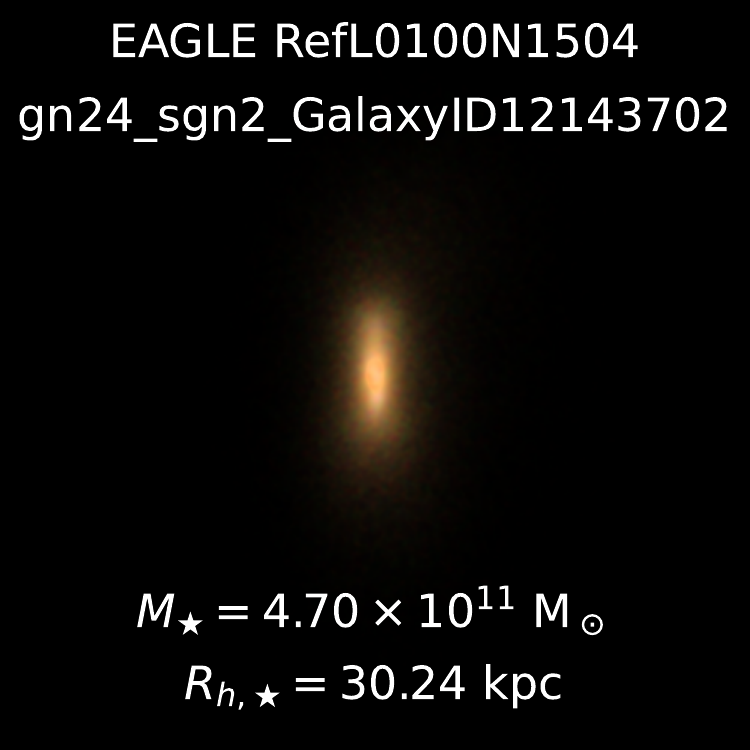}}
\subfigure{\includegraphics[width=0.23\textwidth]{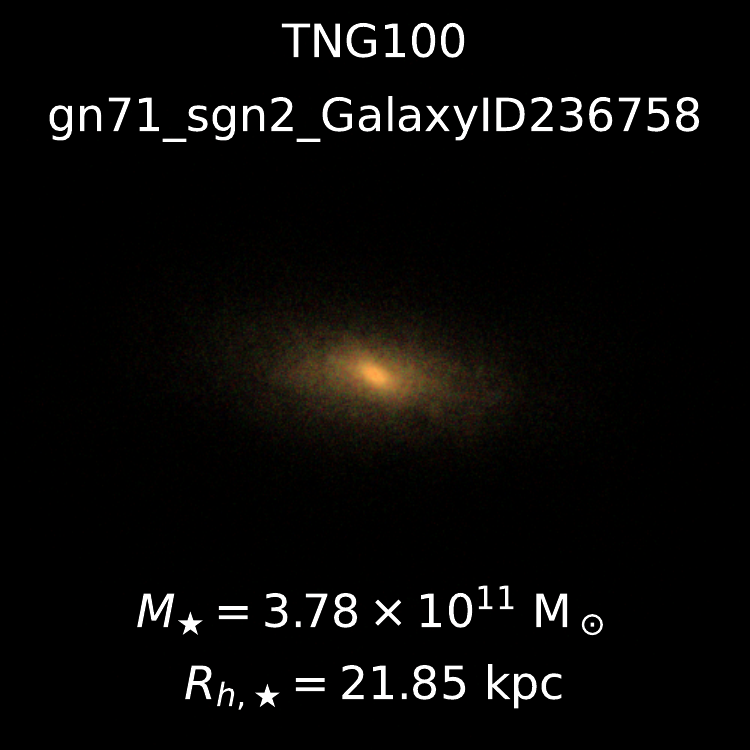}}
\subfigure{\includegraphics[width=0.23\textwidth]{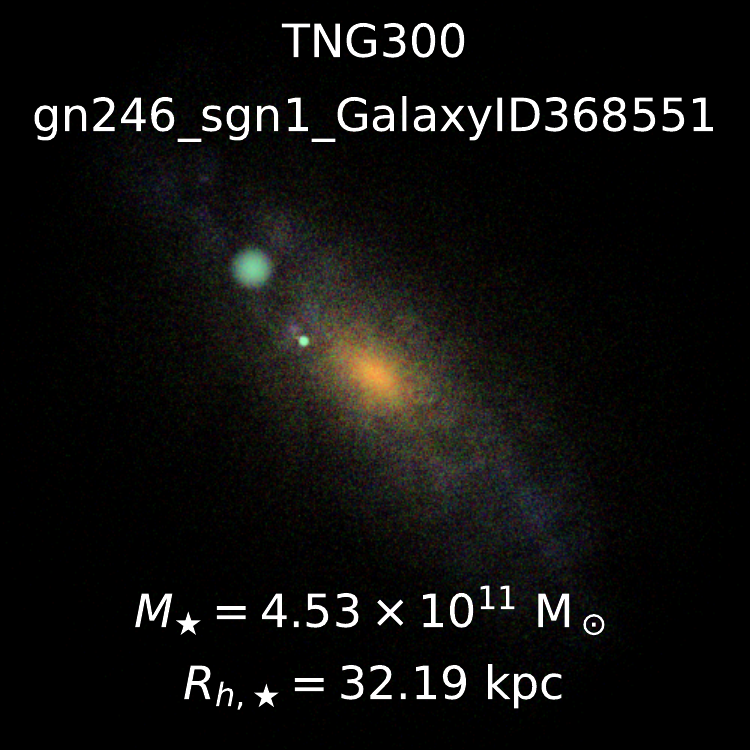}}
\subfigure{\includegraphics[width=0.23\textwidth]{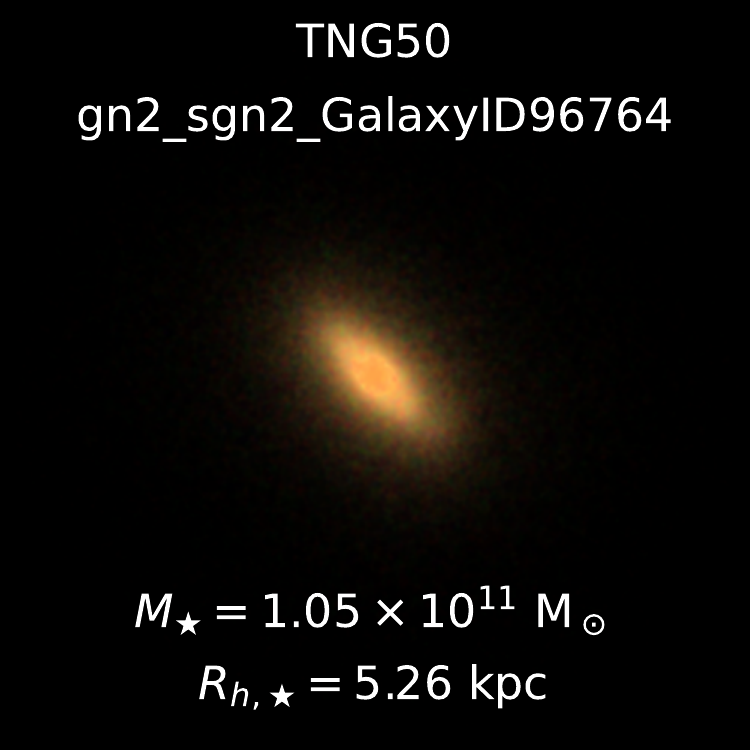}}
\caption{ 
Top panels: mock images of the central galaxies corresponding to our best Fornax-like clusters. From left to right: central galaxies from the best Fornax-like candidates extracted from the \eagle\ RefL0100N1504, TNG100, TNG300, and TNG50 simulations, respectively. The top labels show the corresponding simulation and the complete identification of each galaxy. The bottom labels list the stellar mass $M_\star$ and stellar half-mass radius $R_{\rm h,\star}$. Bottom panels: same as the top panels, but showing mock images of the second most massive galaxy in the corresponding best Fornax-like cluster.
}
\label{fig:NGC1399-likes_mock_images}
\end{figure*}

\begin{figure*}
\centering
\includegraphics[width=\linewidth]{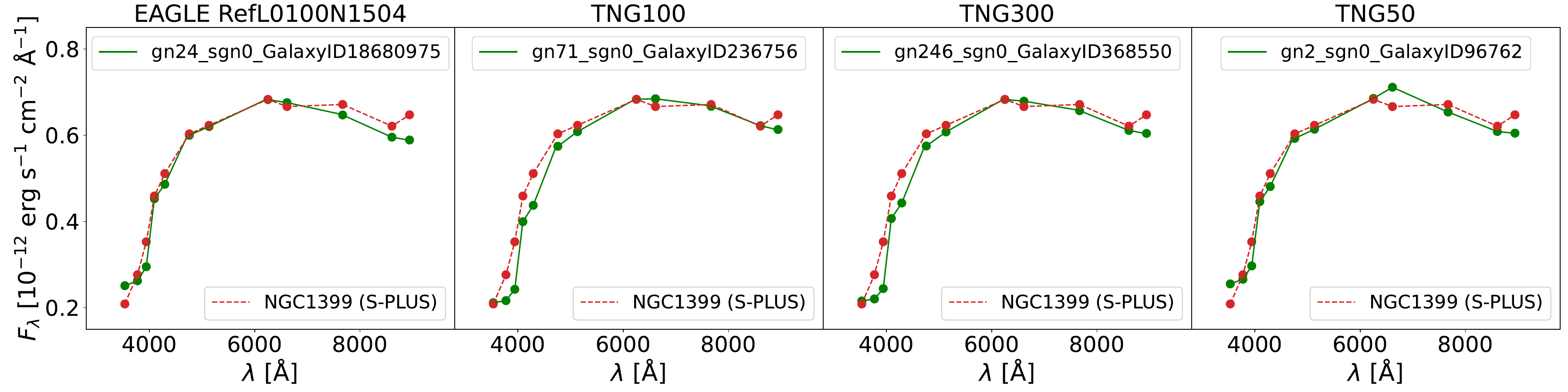}
\caption{
SEDs of the central galaxies from our best Fornax-like candidates. For comparison, the observed SED of NGC\,1399 (the central galaxy of the Fornax cluster) obtained with S-PLUS is shown with red circles and a red dashed line. From left to right, the green circles and green solid lines show the SEDs of the central galaxies corresponding to the best Fornax-like candidates extracted from the \eagle\ RefL0100N1504, TNG100, TNG300, and TNG50 simulations, respectively. The corresponding legend indicates the identification of each galaxy.}
\label{fig:NGC1399-likes_SEDs}
\end{figure*}

Having generated the 12 frames for each galaxy in our Fornax-like clusters, combinations of them can be used to produce RGB images, such as those shown in Fig.~\ref{fig:NGC1399-likes_mock_images}. The top row shows RGB images of the central galaxies in our best Fornax-like clusters (as defined in Sect.~\ref{subsec:Fornax-like_spatial_distribution}), while the bottom row shows the second most massive galaxies in those systems. These images, which provide a first approximation of the products that can be obtained from the \skirt\ outputs, were constructed using the S-PLUS $i$, $r$, and $g$ frames (corresponding to the iSDSS, rSDSS, and gSDSS filters, respectively). The composite images were generated following the scheme of \citet{Lupton2004}, adopting the same stretch and softening parameters for all galaxies. As expected from our selection criteria, the central galaxies in our best Fornax analogs display a spheroidal visual morphology, except for the central galaxy of the best Fornax-like cluster selected from TNG50 (see Sect.~\ref{subsec:fornax_like_clusters}). Although the TNG50 galaxy exhibits a noticeably flatter morphology, it still appears broadly spheroidal.

It is also evident from visual inspection of the mock images that these systems show no clear evidence of disc structures and are dominated by older stellar populations, with no obvious signs of young stars or active star-forming regions. This is consistent with the observed properties of NGC\,1399, the central galaxy of the Fornax cluster. Furthermore, and as imposed by our selection criteria, the central galaxies of our best Fornax-like clusters have stellar masses and stellar half-mass radii consistent with the values estimated for NGC\,1399. In contrast, the second most massive galaxies in our Fornax-like clusters (bottom row) exhibit markedly flattened morphologies. In particular, the galaxy belonging to the $GroupNumber=246$ cluster in TNG300 shows prominent regions of young stars (displayed in blue), together with a clear disc structure. This is especially interesting when compared with NGC\,1316, which observationally is an elongated ellipsoid with extended and prominent dust patches and evidence of a recent merger. It is classified as an SAB0-type lenticular galaxy observed nearly face-on, with an estimated stellar mass in the range $5\times 10^{11} \lesssim M_\star/\rm{M}_\odot \lesssim 8\times10^{11}$ derived from photometry and an appropriate mass-luminosity relation (\citealp{Iodice2017}). Although NGC\,1316 is observed nearly face-on, while the second most massive galaxies in our best simulated Fornax-like clusters appear edge-on, the simulations and observations are broadly consistent in terms of morphology and stellar mass. Modifying the orientation of the simulated observations, which can be done with \skirt, would likely reveal that many of these galaxies are lenticular systems with morphologies similar to that of NGC\,1316.

\subsection{Simulated SEDs}
\label{subsec:fornax_likes_seds}

\begin{figure*}[!ht]
    \centering
    {\includegraphics[width=\linewidth]{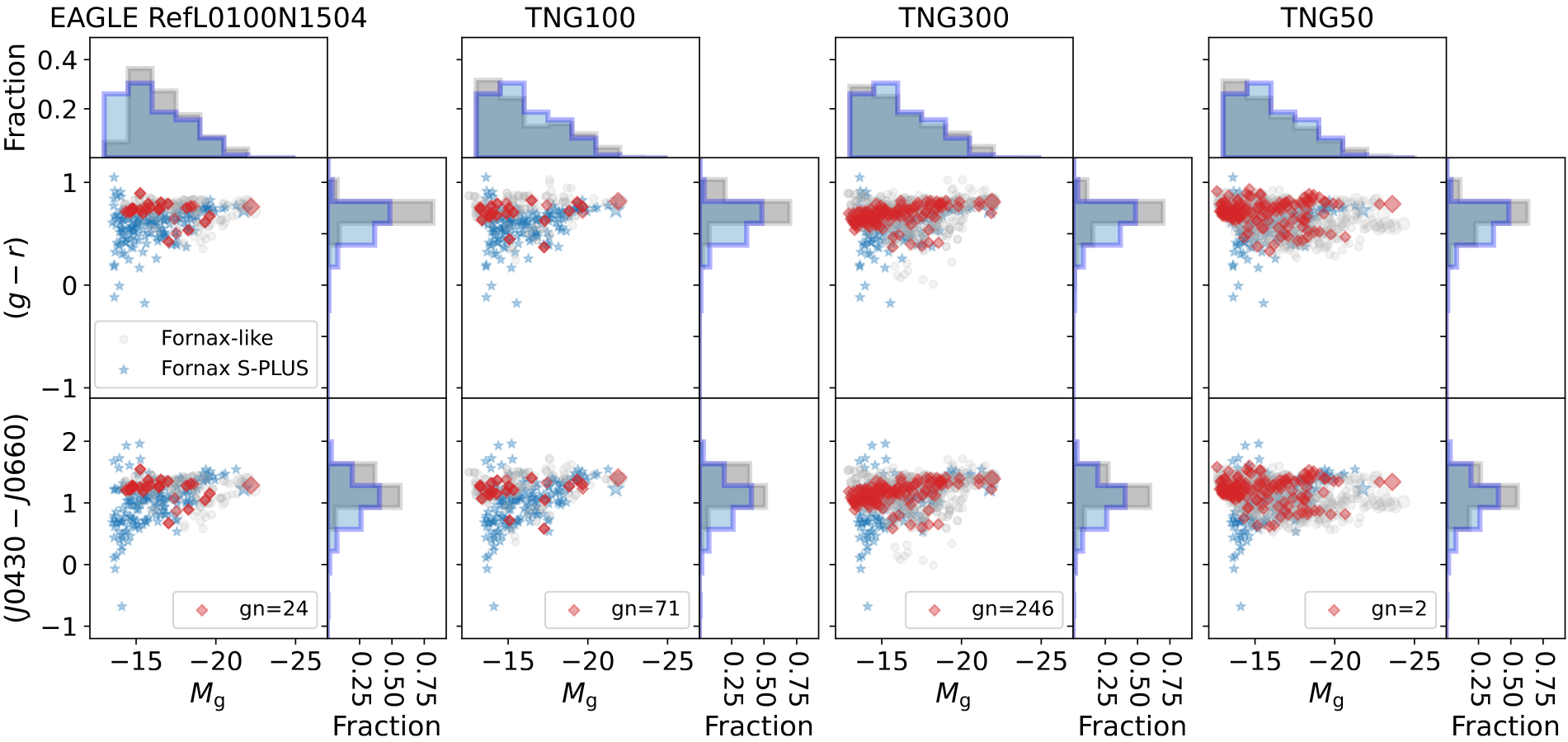}}
    \caption{$g-r$ and $J0430-J0660$ versus $M_{\rm g}$ CMDs of our simulated Fornax-like clusters. From left to right, the panels show Fornax-like clusters from the \eagle\ RefL0100N1504, TNG100, TNG300, and TNG50 simulations (gray circles). Red diamonds represent galaxies in the best Fornax-like cluster defined in this work within each simulation, identified by its $GroupNumber$ ($gn$). For comparison, the corresponding CMD of the Fornax cluster observed with S-PLUS is shown with blue stars. Larger symbols represent central galaxies. The top row and right column panels show, respectively, the distributions (histograms) of magnitudes and colors, with gray (blue) bars corresponding to simulated (observed) galaxies.
    }
    \label{fig:CMDs_best_fornax_likes}
\end{figure*}

Using again the \skirt\ code, we generated the spectral energy distribution (SED) of every galaxy in our Fornax-like clusters, using the 12 S-PLUS bands as a wavelength grid with their corresponding transmission curves{\footnote{Strictly speaking, in this work we generated a sampling of the SEDs using the S-PLUS photometric bands, not the full energy distribution. From now on, SED will refer to the aforementioned sampling.}}. The code generates the SEDs of the simulated systems by calculating the effective wavelength of each filter and assigning to each one its corresponding flux. The simulated spectra obtained with \skirt\ can be directly compared with those observed by the S-PLUS survey.

As an example, Fig.~\ref{fig:NGC1399-likes_SEDs} shows the simulated spectrum of the central galaxy in our best Fornax-like clusters (green solid line), compared to the corresponding SED observed by S-PLUS for NGC\,1399 (red dashed line). Since we are interested in comparing the shape of the simulated SEDs with the observations, rather than their flux values, the spectra shown are normalized at the wavelength $\lambda=6258~{\text{\AA}}$ (S-PLUS $r$ band) with respect to that of NGC\,1399. This normalization was carried out as follows: let $f_{\lambda}$ be the simulated flux (obtained directly with \skirt), $f_{\lambda,\ \rm{norm}}$ the normalized simulated flux (i.e., the one shown with green circles in Fig.~\ref{fig:NGC1399-likes_SEDs}), and $f_{6258~{\text{\AA}},\ \rm{NGC1399}}$ the observed flux of NGC\,1399 at $\lambda=6258~{\text{\AA}}$. The normalized flux at the wavelength $\lambda$ results:

\begin{equation}
    f_{\lambda,\ \rm{norm}}=6258~{\text{\AA}}\left(\dfrac{f_{\lambda}}{\lambda}\right) \left(\dfrac{f_{6258~{\text{\AA}},\ \rm{NGC1399}}}{f_{6258~{\text{\AA}}}}\right)\ .
\end{equation}

If the fluxes of the simulated galaxies are not normalized, minor numerical discrepancies between the simulated and observed SEDs are expected. From this, it can be seen that these simulated spectra are consistent with the S-PLUS observations of NGC\,1399. Although only a few simulated SEDs are shown here as examples, it is important to remember that, using \skirt, we generated the spectra of all galaxies belonging to all our simulated Fornax-like clusters, always using the 12 S-PLUS bands.

\subsection{Simulated magnitudes and colors}
\label{subsec:fornax_likes_magnitudes}

Using the simulated spectra generated with \skirt, we can estimate the apparent magnitudes of every galaxy in our simulated Fornax-like clusters in the 12 S-PLUS photometric bands by convolving the flux obtained from the SEDs with the corresponding transmission curve. In this way, photometric information about the simulated systems (i.e., 12 apparent magnitudes that can be transformed into absolute magnitudes, and all possible combinations of these to derive different colors) can be obtained for a direct comparison with the S-PLUS observations.

As an example, Fig.~\ref{fig:CMDs_best_fornax_likes} shows the color-magnitude diagrams (CMDs) using the broad-band $g-r$ and narrow-band $J0430-J0660$ colors versus the $M_{\rm g}$ magnitude. From left to right, the CMDs correspond to the \eagle\ RefL0100N1504, TNG100, TNG300, and TNG50 simulations. Galaxies in the Fornax-like clusters from the corresponding simulation are represented by gray circles. Red diamonds correspond to galaxies in the best Fornax-like clusters previously defined within each simulation. For comparison, blue stars represent S-PLUS observations of Fornax. Larger symbols correspond to central galaxies. The distributions (histograms) of the corresponding magnitudes and colors are also shown (top row and right column of panels, respectively), indicating the fraction of galaxies in each histogram bin.

\begin{table*}[!ht]
\centering
\caption{Quantitative comparison between simulated and observed $g-r$ colors.}
\small
{
\setlength{\tabcolsep}{8.8pt}
\renewcommand{\arraystretch}{1.} 
\begin{tabular}{ccccccc}
\hline
\hline
 Simulation & $\mu_{\rm b}$ & $\sigma_{\rm b}$ & $\mu_{\rm r}$ & $\sigma_{\rm r}$ & $a_{\rm r}$ & $f_{\rm b}$ \\
    \hline
    EAGLE RefL0100N1504 & 0.38 & 0.01 & 0.71 & 0.08 & $-0.010 \pm 0.002$ & 0.08 \\
    TNG100 & 0.35 & 0.03 & 0.70 & 0.09 & $-0.010 \pm 0.002$ & 0.06 \\
    TNG300 & 0.40 & 0.02 & 0.69 & 0.09 & $-0.011 \pm 0.002$ & 0.08 \\
    TNG50 & 0.47 & 0.05 & 0.73 & 0.08 & $-0.011 \pm 0.002$ & 0.14 \\
    \hline
    Fornax (S-PLUS observations) & 0.42 & 0.07 & 0.65 & 0.09 & $-0.010 \pm 0.002$ & 0.09 \\ 
\hline
\end{tabular}
\tablefoot{The first four rows correspond to our best simulated Fornax-like clusters, while the bottom row corresponds to the S-PLUS observations of the Fornax cluster. From left to right, the columns list the simulation to which the best Fornax-like clusters belong (or the observations, in the bottom row), the mean of the Gaussian fit to the blue galaxy population ($\mu_{\rm b}$), its standard deviation ($\sigma_{\rm b}$), the mean of the Gaussian fit to the red galaxy population ($\mu_{\rm r}$), its standard deviation ($\sigma_{\rm r}$), the slope of the red sequence in the $g-r$ versus $M_{\rm g}$ plane ($a_{\rm r}$), and the fraction of blue galaxies in the cluster ($f_{\rm b}$).}
}
\label{tab:CMR_fit}
\end{table*}

For a better comparison, the observed magnitudes were limited to Fornax galaxies with $M_{\rm g} \leqslant -13.5~\rm{mag}$; i.e., the faintest galaxies (dwarf and low surface brightness galaxies) were discarded from the observed CMDs. Although our Fornax-like clusters contain considerably fewer galaxies than the Fornax cluster (except perhaps in the TNG300 and TNG50 simulations), the simulated magnitudes and colors are consistent with the S-PLUS observations. In particular, the simulated CMDs trace a well-populated and clearly defined red sequence of galaxies, a typical feature in CMDs of galaxy clusters. However, the blue cloud in the simulated CMDs shown here is very sparse and underpopulated compared to the Fornax observations, thus predicting fewer blue galaxies in our Fornax-like clusters. The simulated CMDs also show an excess of bright and blue galaxies (particularly in the TNG300 and TNG50 simulations) compared to the S-PLUS observations of Fornax. Despite this, the agreement between observations and simulations is reasonably good.

Considering the magnitude distributions, the histograms generally show the same behavior for observations and simulations, although there seem to be fewer (more) faint (bright) simulated galaxies than observed ones. The lower number of faint galaxies in the simulated CMDs is mainly due to the restriction imposed on the stellar masses of the simulated galaxies (see Sect.~\ref{subsec:fornax_like_clusters}). Finally, the simulated and observed color distributions show similar behavior in the vicinity of the red sequence (there are more simulated red galaxies there, mainly because our best Fornax-like clusters overlap), again showing the lower number of simulated blue galaxies. Since the simulated magnitudes were generated without taking dust effects into account, the S-PLUS magnitudes and colors (extracted from the S-PLUS DR3 catalogs) had to be corrected for extinction and reddening effects in order to allow a direct comparison. The S-PLUS magnitudes were measured using several apertures optimized for different applications. Here, apertures labeled ``AUTO'' were used, defined in terms of the Kron elliptical aperture to integrate the full flux of extended sources, these magnitudes being the most appropriate for bright objects (see \citealt{AlmeidaFernandes2022} for more details).

Although a detailed statistical analysis of our Fornax-like clusters is beyond the scope of this paper, we performed a simple statistical test to compare simulations and observations. We fitted a double Gaussian to the $g-r$ color distributions of our best Fornax-like clusters and the S-PLUS observations in order to identify the red and blue galaxy populations. We defined the red sequence as galaxies with $g-r$ within $\mu_{\rm r} \pm 2\sigma_{\rm r}$, where $\mu_{\rm r}$ and $\sigma_{\rm r}$ are the mean and standard deviation of the red Gaussian. Similarly, the blue population was defined by $\mu_{\rm b} \pm 2\sigma_{\rm b}$ from the blue Gaussian. The resulting $\mu_{\rm b}$, $\sigma_{\rm b}$, $\mu_{\rm r}$, and $\sigma_{\rm r}$ values for simulations and observations are listed in Table~\ref{tab:CMR_fit}. As can be seen, the simulated red sequences are consistent with the observed in Fornax. The simulated blue populations are also broadly consistent with the observations, although this result should be treated with caution because the observed blue population corresponds to the blue cloud, which is underpopulated in the simulations and includes some bright blue galaxies that do not belong to the blue cloud.

Using the galaxies in the previously defined red populations, we computed the slopes of the simulated and observed red sequences in the $g-r$ vs. $M_{\rm g}$ plane ($a_{\rm r}$) using least-squares linear fits. The slope values obtained are listed in Table~\ref{tab:CMR_fit}. The agreement between the simulated and observed red sequence slopes is remarkably good, at least at first order. Using the computed blue galaxy populations, we also estimated the blue galaxy fractions in our best Fornax-like clusters and in the Fornax observations ($f_{\rm b}$), whose values are also listed in Table~\ref{tab:CMR_fit}. Except for TNG50, the simulated fractions are consistent with the observations. The blue galaxy fractions in our best Fornax-like clusters will be analyzed in detail in a future work.

%--------------------------------------------------------------------

\section{Discussion}
\label{sec:discussion}

We presented here a sample of simulated Fornax-like clusters extracted from cosmological hydrodynamical simulations. We showed some representative examples of such clusters and their corresponding simulated observations, aiming at a direct comparison between simulated and S-PLUS observed quantities in the Fornax cluster. Usually, however, when selecting simulated FoF groups similar to a given galaxy cluster, the selection in the literature is mainly based only on cluster properties. The study of Fornax-like clusters is not an exception, and such systems have been selected using only their virial mass (e.g. \citealp{Venhola2022}; \citealp{Lara-Lopez2022}; \citealp{Ding2023}; \citealp{Chaturvedi2024}), or using both their virial mass and virial radius (e.g. \citealp{Galan2022}). Such works clearly show consistency between the properties of simulated FoF group galaxies and the Fornax observations, but if the goal is a direct comparison, it is important to ensure that the simulated systems reasonably resemble what is observed. In this sense, a simulated FoF group could be selected using only its $M_{\rm vir}$, but when analyzing its central galaxy, for instance, it could be morphologically or kinematically different from NGC\,1399, the central galaxy of Fornax, or could even have a very different stellar mass and size. Hence, from an observational point of view, such a simulated system could be labeled as a not-so-good Fornax-like candidate. To avoid such conflicts, in this work we carried out a more refined selection of Fornax-like candidates compared to previous studies, using not only the virial mass of the simulated clusters, but also properties of their central galaxy (namely stellar mass, half-mass radius, and visual morphology), in order to select simulated FoF groups that, at least in their inner regions and in terms of their main galaxy, can be considered reasonably good Fornax-like clusters based on what is directly observed in Fornax. The results presented throughout this work suggest that our selection criteria lead to simulated features and properties that are consistent with the Fornax observations. Note that if we did not apply our morphology constraint to the central galaxies, the number of simulated clusters would increase, particularly in the TNG300 simulation: without such a constraint, there would be 45, 24, and 1097 simulated clusters from, respectively, {\sc{eagle}} RefL0100N1504, TNG100, and TNG300. However, as stated previously, systems with a non-spherical central galaxy could observationally be considered poor Fornax-like candidates.

\begin{figure*}
\centering
\includegraphics[width=0.9\textwidth]{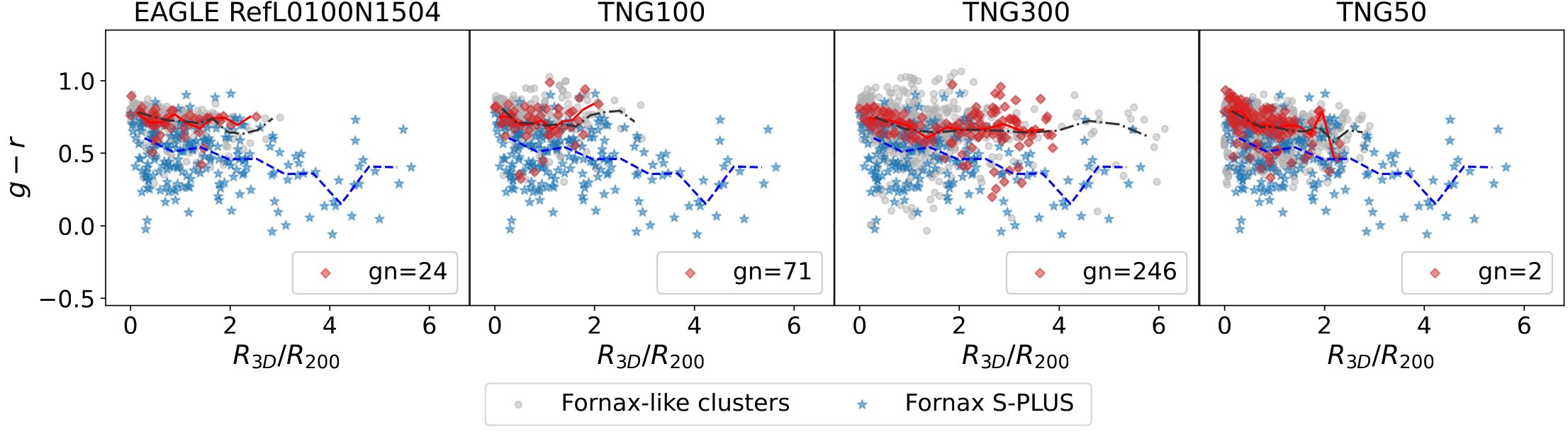}
    \caption{$g-r$ color as a function of the 3D cluster-centric distance normalized by the virial radius ($R_{\rm 3D}/R_{200}$). Gray circles represent galaxies in our simulated Fornax-like clusters. From left to right, the panels show systems extracted from the \eagle\ RefL0100N1504, TNG100, TNG300, and TNG50 simulations. In each panel, red diamonds correspond to galaxies in the best Fornax-like cluster defined within each simulation, identified by its $GroupNumber$ ($gn$). Blue stars represent Fornax galaxies observed with S-PLUS. The dash-dotted black, solid red, and dashed blue lines depict the median $g-r$ versus $R_{\rm 3D}$ relation for all the corresponding Fornax-like clusters, the best Fornax-like cluster, and the observed Fornax cluster, respectively.}
    \label{fig:gr_vs_R}
\end{figure*}

Another important aspect in assessing whether a simulated FoF group is a good Fornax-like candidate is its galaxy spatial distribution. The similarity between our Fornax-like groups and the Fornax cluster may depend on the simulation box size and resolution. In \eagle\ RefL0100N1504, TNG100, and TNG50, the smaller box sizes make Fornax-like structures less common, explaining the limited number of extracted candidates. In addition, their resolution may cause the FoF algorithm to separate two large substructures, analogous to the central Fornax region and Fornax A, into distinct halos with different $GroupNumber$ values, preventing the recovery of a Fornax-like galaxy distribution. In contrast, the larger but lower-resolution TNG300 simulation is more likely to contain massive halos composed of multiple important substructures, such as our best Fornax-like candidate from this run. In this work, groups were selected solely by their $GroupNumber$, considering only individual FoF groups that satisfy our criteria. A future analysis combining distinct FoF groups into systems with Fornax-like galaxy distributions will broaden our catalog of Fornax analogs and allow a more detailed study of central Fornax-like and Fornax A-like regions.

Regarding the dust-free \skirt\ runs performed in this work, neglecting dust may appear to be a strong assumption, particularly since the comparison with observations involves several optical bands, including narrow bands sensitive to emission lines. However, FIR observations indicate that only a small fraction of Fornax galaxies ($\sim 13\%$) contain detectable dust (e.g., \citealp{Fuller2014, Zabel2019}). In addition, extinction estimates for $\sim 200$ Fornax galaxies show that most have optical depths $\tau_{\rm ISM}<0.4$ (Thainá-Batista et al., in prep.). Considering this, together with the exponentially higher computational cost of including dust in {\sc{skirt}}, we modeled our mock observations without dust. A detailed analysis of dust effects and their impact on simulated colors will be presented in a future work.

With respect to the simulated images generated with \skirt, it is important to note that the ones presented here serve only as examples, and data cubes were generated in the 12 S-PLUS bands for all galaxies in all our Fornax-like clusters. Any of the frames in the data cubes can be combined to obtain RGB images, and, with some care, all of them can be combined for a given galaxy. Moreover, different frames can be used to highlight specific features in the simulated images, such as active or recent star-forming regions, disc structure, spiral arms, interactions, and merger remnants. In order to carry out a more robust study of these images through direct comparison with observations, it is necessary to add certain realism to them, such as background noise, statistical errors related to observational measurements and instrumental noise, and the convolution of the images with a point spread function (PSF), among other typical observational effects that must be taken into account in galaxy observations. Furthermore, what is shown here represents only a first approximation to the generation of simulated galaxy images using \skirt, and the visualizations can be substantially improved in terms of smoothing and separation of stellar populations. This, combined with the possibility of studying visualizations of simulated galaxies in different orientations, enables a rigorous and exhaustive analysis of mock images (e.g., \citealp{Baes2024}).

The frames generated by \skirt\ can also be used for more detailed analyses. Since they contain the surface brightness at each pixel of the image in the 12 S-PLUS bands, they can be used, for example, to construct surface brightness profiles, which can then be appropriately fitted with a Sérsic law (\citealp{Sersic1968}), as well as to construct maps similar to those obtained from integral field spectroscopy, allowing the analysis of stellar populations with spatial resolution within a single galaxy (see, for example, Figure 10 of \citealp{Thaina2023} and \citealp{Thaina2025}). Furthermore, these mock images can be used to estimate several galaxy structural parameters (e.g., inclination, ellipticity, and semi-axes), in order to perform a comprehensive analysis of parametric morphologies. Hence, a robust morphological classification of simulated galaxies based on the obtained images can be carried out, complementing quantitative, visual, and morpho-kinematic analyses (e.g., \citealp{RodriguezGomez2019,Jang2023,Martin2025}). These topics comprise a large number of analyses that can be performed using the images generated with \skirt, and their detailed study will be presented in future works, since the main goal of this paper is to introduce our sample of Fornax-like clusters, as well as to describe the methodology followed to compute simulated observations, showing here a first comparison between them and the S-PLUS observations.

The most important mock observations obtained in this work comprise the computation of SEDs and photometric magnitudes and colors using the S-PLUS filter system by means of the \skirt\ code. For instance, using the \eagle\ and \tng\ simulations, mock SEDs and synthetic photometry have been computed in previous works (e.g., \citealp{Trayford2017,RodriguezGomez2019,Trcka2022,Gebek2024}), and magnitudes in the SDSS filter system can be queried from the simulation databases. These include the broad-band filters of S-PLUS, but not the S-PLUS narrow-band filters. In this sense, the implementation of the 12 S-PLUS photometric bands was one of the most important steps in this work, since such data were not directly available from the simulations used. This opens numerous opportunities to perform rigorous comparisons between simulated and observed galaxies and constitutes a valuable contribution to the S-PLUS collaboration and to the study of the Fornax cluster region.

We computed the SEDs for every galaxy in our simulated Fornax-like clusters, showing in particular the SEDs of the central galaxies corresponding to our best Fornax-like clusters. The agreement between these and the observed SED of NGC,1399 is remarkably good, which clearly highlights the success of the technique developed in this work to construct the SEDs with \skirt. With our computed spectra, it is possible, for example, to statistically analyze the SEDs of a given galaxy population in our Fornax-like clusters. Such populations could be defined, for instance, as a function of galaxy morphology, stellar mass, chemical properties, or star formation rate, among other properties available in the simulations. This statistical analysis of simulated SEDs, as well as their comprehensive comparison with SEDs observed by S-PLUS, is left for future work.

With respect to the simulated colors and magnitudes of galaxies in our Fornax-like clusters, although they are consistent with the S-PLUS observations of Fornax, there are differences between simulations and observations in the CMDs that cannot be ignored, namely the underpopulation of the simulated blue cloud and the excess of simulated bright blue galaxies with respect to the observations. The lack of galaxies in the simulated blue cloud is, in part, a selection effect: as we verified, if the simulated clusters are selected using only their $M_{\rm vir}$ and $R_{\rm vir}$, the blue cloud appears relatively well populated, but as the properties of the central galaxies are restricted to satisfy our final selection criteria, the number of galaxies comprising the blue cloud decreases considerably. In addition, the underpopulation of the simulated blue clouds may be related to numerical effects, likely due to stronger quenching mechanisms in the simulations compared to the actual properties of the cluster: models of stellar feedback and/or environmental effects in the simulations may be more efficient than they should be in low-brightness galaxies, excessively limiting their star formation activity (see, e.g., \citealp{Schaye2015} and \citealp{Nelson18} for the \eagle\ and \tng\ simulations, respectively). On the other hand, the excess of bright blue simulated galaxies may be caused by an overestimation of star formation rates in such galaxies, related to an insufficient suppression of star formation activity due to AGN feedback (e.g., \citealp{Trayford2017}; \citealp{Weinberger2018}). Hence, the differences between the simulated and observed CMDs may be mainly related to feedback modeling, but this cannot be assessed within the scope of this work. It is worth noting that the different CMDs shown in this work are only examples that allow us to assess the consistency between the magnitudes observed by S-PLUS in Fornax and those estimated using \skirt\ for the galaxies in our Fornax-like clusters. Many additional combinations of magnitudes, especially colors, can provide valuable information about galaxy stellar populations (such as ages, metallicities, and star formation rates), although a multi-wavelength approach would be ideal (e.g., \citealp{Humire2025}).

The results presented here for galaxies in our Fornax-like clusters are mainly based on quantities that can be directly observed in galaxy surveys, particularly in the S-PLUS survey. All these mock observations were computed using the \skirt\ code, configuring the simulated instruments in a way similar to the S-PLUS survey. However, the simulation databases contain many additional galaxy properties that can be studied. Therefore, we can carry out detailed analyses of several galaxy properties, some of which can also be derived from observations (specifically from simulated observations). In particular, SEDs, magnitudes, and colors can be used to estimate physical galaxy properties such as stellar masses, metallicities, and star formation rates. These simulated quantities can be extensively used in future analyses, including morpho-kinematic diagnostics, chemical and formation histories, and the measurement of the large-scale conformity signal traced and produced by galaxies in our Fornax-like systems (e.g., \citealp{Palma2025}).

The simulations used to extract our Fornax-like clusters also allow us to analyze properties (both observable quantities computed from our implementation of \skirt\ and physical parameters extracted directly from the simulation databases, as well as physical properties derived from our simulated observations) as a function of cluster-centric distance. Such a study is fundamental for determining the physical processes involved in the formation histories of galaxies in different regions of our Fornax-like systems, and will certainly help propose possible formation and evolution scenarios for the Fornax cluster. As a simple example, Fig.~\ref{fig:gr_vs_R} shows the $g-r$ color as a function of the 3D cluster-centric distance ($R_{\rm 3D}/R_{200}$) for our Fornax-like clusters. From left to right, the panels correspond to systems extracted from \eagle\ RefL0100N1504, TNG100, TNG300, and TNG50. Gray circles represent galaxies in all our Fornax-like systems, while red diamonds depict galaxies in the corresponding best Fornax-like cluster within each simulation, as defined in Section~\ref{subsec:fornax_like_clusters} (see also Table~\ref{tab:best_Fornax-likes}). For comparison, blue stars correspond to S-PLUS observations of Fornax. It is important to note that, in the observations, S-PLUS data only allow us to compute projected distances rather than 3D distances. The dash-dotted black, solid red, and dashed blue lines represent the median $g-r$ as a function of cluster-centric distance for all the corresponding Fornax-like clusters, the respective best Fornax-like candidate, and the S-PLUS observations of Fornax, respectively. In general terms, the median simulated $g-r$ colors for our Fornax-like clusters are $\sim0.2~\rm{mag}$ redder than the observed values at a fixed location within the cluster, although there is considerable dispersion in both simulations and observations. Nevertheless, the trends in the simulated and observed median $g-r$ versus $R$ relations are the same: galaxies tend to become bluer at larger cluster-centric distances. In the case of our best simulated Fornax-like cluster from TNG300 (which is our strongest Fornax-like candidate, as stated previously), this trend remains valid at least up to $R_{\rm 3D}\sim 1.5~R_{200}$, following an almost linear relation. However, at $R_{\rm 3D}\gtrsim1.5~R_{200}$, its median $g-r$ appears to become approximately constant and independent of $R_{\rm 3D}$, with a clearly increasing dispersion in $g-r$ and the presence of several blue galaxies. This change in the behavior of $g-r$ as a function of cluster-centric distance is likely due to changing environmental conditions within the system (i.e., the transition from the central NGC,1399-like region to the outer NGC,1316-like region), and is probably driven by physical processes that strongly depend on the environment.

%--------------------------------------------------------------------

\section{Concluding remarks}
\label{sec:conclusions}

In this work, we presented a catalog of simulated Fornax-like clusters extracted from the state-of-the-art \eagle\ and \tng\ simulations. Our selection of Fornax-like clusters was based on the virial mass estimated for the Fornax cluster, along with observed properties (stellar mass, size of the stellar body, and visual morphology) of its central galaxy. By implementing the {\sc{skirt}} radiative transfer code, we simulated the stellar emission of every galaxy in our Fornax-like clusters, generating mock observations of such systems. We obtained synthetic images (Sect.~\ref{subsec:fornax_likes_mock_images}), spectra (Sect.~\ref{subsec:fornax_likes_seds}), and photometric magnitudes (Sect.~\ref{subsec:fornax_likes_magnitudes}) for all galaxies in our Fornax-like clusters, using the 12 S-PLUS photometric bands and their corresponding transmission curves, in order to obtain quantities that can be directly compared with those observed by the S-PLUS survey.

From our complete sample of Fornax-like clusters, we selected four systems (one from each simulation used) based on the spatial distribution of their galaxies, particularly taking into account the distance between the central galaxy and the second most massive galaxy. These four clusters were defined here as our best Fornax-like clusters. We present here our first steps toward a comprehensive analysis of simulated observations of such systems and their detailed comparison with S-PLUS observations of the Fornax cluster. In summary, the preliminary results shown in this work suggest that our selection criteria are reasonably adequate, with the simulated observations obtained using \skirt\ being consistent with observational results from the S-PLUS survey.

The simulated SEDs computed with \skirt\ can be directly compared with observed ones. In particular, the four SEDs corresponding to the central galaxies of our best Fornax-like clusters are consistent with the SED of NGC\,1399 (Fig.~\ref{fig:NGC1399-likes_SEDs}). Using the spectra obtained with \skirt, we calculated magnitudes and colors for all galaxies in our Fornax-like clusters (Sect.~\ref{subsec:fornax_likes_magnitudes}) in the 12 S-PLUS bands. These simulated quantities can be directly compared with those obtained observationally from the survey. Furthermore, by having magnitudes in the 12 S-PLUS bands, all simulated color combinations can be derived, making it possible to construct different color--magnitude diagrams (CMDs) that can be directly compared with the observed ones. In particular, the CMDs of our best simulated Fornax-like clusters are consistent with the S-PLUS observations, despite some differences that we identified (Fig.~\ref{fig:CMDs_best_fornax_likes}). Nevertheless, the simulated CMDs show features typically found in observations, indicating that our computation of synthetic photometric quantities using \skirt\ is reasonably good. The S-PLUS magnitudes of galaxies in our Fornax-like clusters are not directly available in the simulation databases, which further highlights the importance of our method.

The analysis carried out so far and presented in this work is, of course, open to significant extensions, modifications, and improvements, especially if a complete comparison between simulations and observations is desired. On the other hand, the simulated quantities derived in this work can be used, for example, to estimate galaxy masses and metallicities in our Fornax-like clusters using methods similar to those implemented in spectro-photometric analyses. Thus, simulated properties derived from our mock observations can be directly and comprehensively compared with real observations. The results presented here represent our first steps toward this objective. Among the future works we plan to carry out, the most important include improving the generation of simulated galaxy images in our Fornax-like clusters (aiming at a robust morphological classification of galaxies), analyzing stellar populations with spatial resolution for statistical galaxy samples, performing as exhaustive and comprehensive a comparison as possible between simulations and S-PLUS observations, and deriving from simulated observations (if possible) physical quantities of simulated galaxies that can be compared with those derived from S-PLUS observations. Finally, and as our main goal, the evolution histories of galaxies in our Fornax-like clusters will be analyzed through the simulations, as well as the evolution and assembly history of each cluster as a whole. This will clearly allow us to establish formation and evolution scenarios for the Fornax cluster within its rich environment, connecting them with the large-scale structure scenario, and to investigate possible mechanisms responsible for quenching the star formation activity of Fornax cluster members.

It is clear that, from a numerical point of view, there are many topics and analyses to address within the framework of the S+FP. In particular, our spectro-photometric simulated observations and mock images of galaxies in Fornax-like clusters provide a very rich and powerful data source for comparison with S-PLUS observations and for testing formation and evolution scenarios of the Fornax cluster. Furthermore, given the complex interactions among the diverse physical processes that affect galaxy evolution in general (e.g., star formation, chemical enrichment, different feedback processes, matter accretion and ejection, etc.), and particularly in dense environments (e.g., interactions, mergers, ram-pressure stripping, etc.), current hydrodynamical cosmological simulations constitute a fundamental tool for studying possible formation scenarios of observed systems. Testing and refining models that describe the different processes involved, as well as carrying out detailed comparisons between simulated and observed properties, will undoubtedly help determine how accurately current simulations reproduce both simple and complex features in such environments, and, in particular, test different hypotheses regarding the past, present, and future of the Fornax cluster. With our results and methodology, we expect to carry out a comprehensive, robust, and self-consistent study of Fornax-like clusters within the S+FP, as well as a complete comparison between simulations and S-PLUS observations of Fornax. Achieving this will clearly be of interest to both observational and computational astronomy, contributing to the synergy between them.

%--------------------------------------------------------------------

\begin{acknowledgements}

LJZ, AVSC, RFH and ARL acknowledge financial support from Consejo Nacional de Investigaciones Científicas y Técnicas (CONICET) (PIP 1504), Agencia I+D+i (PICT 2019–03299) and Universidad Nacional de La Plata (Argentina). AVSC also thanks Fundação de Amparo à Pesquisa do Estado de São Paulo (FAPESP) for the support grant $2025/05085-1$. MEDR acknowledges support from {\it Agencia Nacional de Promoci\'on de la Investigaci\'on, el Desarrollo Tecnol\'ogico y la Innovaci\'on} (Agencia I+D+i, PICT-2021-GRF-TI-00290, Argentina). 
DP acknowledges financial support from ANID through FONDECYT Postdoctorado Project 3230379. DP also acknowledges support from the Agencia Nacional de Investigación y Desarrollo (ANID) through the Millennium Science Initiative Program NCN2024\_112.  MCA acknowledges financial support from FONDECYT Iniciación 11240540 and ANID BASAL project FB210003. DP acknowledges the financial support from Fundaçao de Amparo à Pesquisa do Estado de São Paulo (FAPESP; project 2011/51680-6).  RFH also thanks CAPES for financial support under the program Move La America 2025. ARL also acknowledges the grant 2025/09544-0 from São Paulo Research Foundation (FAPESP). PKH gratefully acknowledges the Fundação de Amparo à Pesquisa do Estado de São Paulo (FAPESP) for the support grant 2023/14272-4. 
We acknowledge the Virgo Consortium for making their
simulation data available. The EAGLE simulations were performed
using the DiRAC-2 facility at Durham, managed by the ICC, and the
PRACE facility Curie based in France at TGCC, CEA, Bruy\`eres-le-
Ch\^atel. We acknowledge the IllustrisTNG team for making their simulation data available. The IllustrisTNG simulations were undertaken with compute time awarded by the Gauss Centre for Supercomputing (GCS) under GCS Large-Scale Projects GCS-ILLU and GCS-DWAR on the GCS share of the supercomputer Hazel Hen at the High Performance Computing Center Stuttgart (HLRS), as well as on the machines of the Max Planck Computing and Data Facility (MPCDF) in Garching, Germany. The S-PLUS project, including the T80-South robotic telescope and the S-PLUS scientific survey, was founded as a partnership between the Fundação de Amparo à Pesquisa do Estado de São Paulo (FAPESP), the Observatório Nacional (ON), the Federal University of Sergipe (UFS), and the Federal University of Santa Catarina (UFSC), with important financial and practical contributions from other collaborating institutes in Brazil, Chile (Universidad de La Serena), and Spain (Centro de Estudios de Física del Cosmos de Aragón, CEFCA). We further acknowledge financial support from the São Paulo Research Foundation (FAPESP), Fundação de Amparo à Pesquisa do Estado do RS (FAPERGS), the Brazilian National Research Council (CNPq), the Coordination for the Improvement of Higher Education Personnel (CAPES), the Carlos Chagas Filho Rio de Janeiro State Research Foundation (FAPERJ), and the Brazilian Innovation Agency (FINEP). The authors who are members of the S-PLUS collaboration are grateful for the contributions from CTIO staff in helping in the construction, commissioning and maintenance of the T80-South telescope and camera. We are also indebted to Rene Laporte and INPE, as well as Keith Taylor, for their important contributions to the project. From CEFCA, we particularly would like to thank Antonio Marín-Franch for his invaluable contributions in the early phases of the project, David Cristóbal-Hornillos and his team for their help with the installation of the data reduction package jype version 0.9.9, César Íñiguez for providing 2D measurements of the filter transmissions, and all other staff members for their support with various aspects of the project. 
\end{acknowledgements}

%--------------------------------------------------------------------

\bibliographystyle{aa} % style aa.bst
\bibliography{aa58912-26} % your references Yourfile.bib

%--------------------------------------------------------------------

\begin{appendix} 

\section{Full sample of selected Fornax-like clusters}

Table~\ref{tab:clusters_dist_massive_galaxies} shows the full list of our simulated Fornax-like clusters, selected as described in Sect.~\ref{subsec:fornax_like_clusters}. Within the corresponding simulation, each cluster is identified with an unique {\it{GroupNumber}} ($gn$). Each of our Fornax-like clusters is characterized by its virial mass ($M_{\rm 200}$), virial radius ($R_{\rm 200}$), stellar masses of the two most massive galaxies ($M_0$ and $M_{\rm sec}$, with $M_0$ the stellar of the central, most massive galaxy), the 3D distance between these galaxies ($d$), that distance projected onto the three coordinate planes of the simulated cosmic box ($d_{\rm xy}$, $d_{\rm xz}$, and $d_{\rm yz}$), and the number of galaxies $N$ with $M_\star \geqslant 5\times10^8~ \rm{M}_\odot$ within each simulated group.

\begingroup
\definecolor{lightcopper}{rgb}{.93, .76, .58}
\begin{table*}
\centering
\caption{Fornax-like clusters according to the selection criteria adopted in this work.}
\small
{
\setlength{\tabcolsep}{10pt}
\renewcommand{\arraystretch}{1.} 
\vspace{-0.24cm}
\begin{tabular}{cccccccccc}
\hline
\hline
$gn$ & $\log(M_{\rm 200})$ & $R_{\rm 200}$ & $\log(M_{\rm 0})$ & $\log(M_{\rm sec})$ & $d$ & $d_{\rm xy}$ & $d_{\rm xz}$ & $d_{\rm yz}$ & $N$ \\
 & $[\rm{M_\odot}]$ & $[\rm{Mpc}]$ & $[\rm{M_\odot}]$ & $[\rm{M_\odot}]$ & $[\rm{Mpc}]$ & $[\rm{Mpc}]$ & $[\rm{Mpc}]$ & $[\rm{Mpc}]$ & \\
 \hline
\multicolumn{10}{c}{\eagle\ RefL0100N1504} \vspace{0cm}
\\
14 & 13.68 & 0.76 & 11.45 & 11.25 & 0.14 & 0.05 & 0.13 & 0.14 & 75 \\
19 & 13.75 & 0.81 & 11.48 & 11.16 & 0.38 & 0.36 & 0.37 & 0.18 & 49 \\
22 & 13.73 & 0.79 & 11.68 & 10.78 & 0.26 & 0.16 & 0.24 & 0.22 & 41 \\
24 & 13.68 & 0.77 & 11.67 & 10.54 & 0.62 & 0.62 & 0.11 & 0.62 & 50 \\
25 & 13.68 & 0.76 & 11.50 & 11.16 & 0.30 & 0.03 & 0.30 & 0.30 & 37 \\
29 & 13.59 & 0.72 & 11.62 & 10.79 & 0.79 & 0.34 & 0.73 & 0.77 & 30 \\
32 & 13.57 & 0.71 & 11.66 & 10.11 & 0.14 & 0.12 & 0.13 & 0.08 & 30 \\
34 & 13.60 & 0.72 & 11.57 & 10.55 & 0.22 & 0.22 & 0.21 & 0.08 & 31 \\
38 & 13.56 & 0.70 & 11.48 & 10.66 & 0.58 & 0.56 & 0.47 & 0.37 & 37 \\
60 & 13.42 & 0.63 & 11.43 & 10.64 & 0.15 & 0.15 & 0.01 & 0.15 & 24 \\
\hline
\multicolumn{10}{c}{TNG100} \vspace{0cm}\\
47 & 13.76 & 0.82 & 11.60 & 11.33 & 0.44 & 0.25 & 0.41 & 0.39 & 91 \\
71 & 13.39 & 0.61 & 11.58 & 10.72 & 0.78 & 0.74 & 0.57 & 0.59 & 53 \\
74 & 13.38 & 0.61 & 11.48 & 11.25 & 0.32 & 0.28 & 0.32 & 0.17 & 55 \\
84 & 13.28 & 0.56 & 11.57 & 10.57 & 1.03 & 0.40 & 1.03 & 0.96 & 48 \\
93 & 13.31 & 0.58 & 11.41 & 11.39 & 0.24 & 0.19 & 0.16 & 0.22 & 45 \\
122 & 13.25 & 0.55 & 11.63 & 10.21 & 0.09 & 0.08 & 0.09 & 0.04 & 34 \\
\hline
\multicolumn{10}{c}{TNG300} \vspace{0cm} \\
246 & 13.76 & 0.81 & 11.66 & 11.66 & 2.26 & 2.25 & 1.27 & 1.88 & 138 \\
339 & 13.49 & 0.66 & 11.52 & 11.28 & 3.63 & 2.61 & 3.56 & 2.62 & 96 \\
442 & 13.89 & 0.90 & 11.69 & 11.09 & 0.55 & 0.20 & 0.55 & 0.52 & 76 \\
475 & 13.87 & 0.89 & 11.46 & 11.42 & 0.65 & 0.64 & 0.55 & 0.37 & 68 \\
680 & 13.72 & 0.79 & 11.65 & 10.88 & 0.53 & 0.33 & 0.53 & 0.43 & 54 \\
788 & 13.68 & 0.76 & 11.63 & 10.91 & 0.25 & 0.14 & 0.23 & 0.22 & 48 \\
799 & 13.50 & 0.67 & 11.45 & 11.37 & 0.75 & 0.51 & 0.61 & 0.70 & 45 \\
827 & 13.56 & 0.70 & 11.63 & 11.31 & 1.78 & 0.93 & 1.61 & 1.70 & 41 \\
846 & 13.73 & 0.80 & 11.63 & 11.03 & 0.49 & 0.25 & 0.44 & 0.48 & 33 \\
888 & 13.60 & 0.72 & 11.68 & 11.07 & 1.21 & 0.55 & 1.21 & 1.08 & 44 \\
894 & 13.68 & 0.77 & 11.57 & 10.60 & 0.42 & 0.18 & 0.41 & 0.39 & 50 \\
912 & 13.67 & 0.76 & 11.69 & 10.50 & 0.49 & 0.19 & 0.45 & 0.49 & 23 \\
927 & 13.71 & 0.78 & 11.51 & 11.32 & 0.34 & 0.28 & 0.25 & 0.30 & 31 \\
1061 & 13.39 & 0.61 & 11.55 & 10.80 & 1.33 & 1.10 & 1.33 & 0.75 & 49 \\
1097 & 13.53 & 0.68 & 11.42 & 10.83 & 1.21 & 0.99 & 0.81 & 1.14 & 49 \\
1115 & 13.42 & 0.63 & 11.46 & 10.77 & 0.97 & 0.95 & 0.65 & 0.74 & 45 \\
1124 & 13.59 & 0.71 & 11.65 & 10.79 & 0.17 & 0.08 & 0.15 & 0.16 & 30 \\
1125 & 13.58 & 0.71 & 11.67 & 10.72 & 0.37 & 0.09 & 0.36 & 0.37 & 37 \\
1269 & 13.53 & 0.68 & 11.56 & 10.75 & 0.69 & 0.53 & 0.67 & 0.48 & 25 \\
1285 & 13.50 & 0.67 & 11.60 & 10.50 & 0.70 & 0.61 & 0.37 & 0.69 & 32 \\
1286 & 13.56 & 0.70 & 11.65 & 10.77 & 0.58 & 0.52 & 0.28 & 0.57 & 22 \\
1333 & 13.33 & 0.59 & 11.44 & 11.01 & 0.81 & 0.77 & 0.81 & 0.24 & 29 \\
1351 & 13.46 & 0.64 & 11.57 & 10.60 & 0.45 & 0.13 & 0.45 & 0.44 & 22 \\
1439 & 13.44 & 0.63 & 11.41 & 10.70 & 0.13 & 0.07 & 0.12 & 0.12 & 28 \\
1445 & 13.46 & 0.65 & 11.56 & 10.80 & 0.27 & 0.12 & 0.27 & 0.24 & 30 \\
2130 & 13.23 & 0.54 & 11.45 & 10.57 & 0.60 & 0.40 & 0.47 & 0.59 & 19 \\
2156 & 13.34 & 0.59 & 11.42 & 10.80 & 0.47 & 0.41 & 0.24 & 0.46 & 20 \\
2400 & 13.22 & 0.54 & 11.52 & 10.31 & 1.02 & 0.98 & 0.38 & 0.98 & 20 \\
2437 & 13.22 & 0.54 & 11.48 & 10.45 & 0.05 & 0.04 & 0.03 & 0.03 & 15 \\
\hline
\multicolumn{10}{c}{TNG50} \vspace{0cm} \\
2 & 13.81 & 0.85 & 12.22 & 11.02 & 1.03 & 1.00 & 0.25 & 1.02 & 141 \\
3 & 13.55 & 0.69 & 12.08 & 11.56 & 1.27 & 1.05 & 0.85 & 1.18 & 160 \\
4 & 13.51 & 0.67 & 11.51 & 11.47 & 1.60 & 1.59 & 1.55 & 0.45 & 129 \\
5 & 13.33 & 0.58 & 11.74 & 11.67 & 50.42 & 50.42 & 0.59 & 50.42 & 110 \\
6 & 13.54 & 0.69 & 11.90 & 11.20 & 0.82 & 0.44 & 0.79 & 0.73 & 103 \\
11 & 13.49 & 0.66 & 11.74 & 11.49 & 0.40 & 0.24 & 0.34 & 0.38 & 54 \\
13 & 13.39 & 0.61 & 11.48 & 11.12 & 0.09 & 0.07 & 0.09 & 0.07 & 47 \\
15 & 13.24 & 0.55 & 11.78 & 10.59 & 0.59 & 0.45 & 0.40 & 0.59 & 45 \\
17 & 13.22 & 0.54 & 11.77 & 10.71 & 0.76 & 0.74 & 0.27 & 0.74 & 27 \\
18 & 13.14 & 0.51 & 11.80 & 11.12 & 0.12 & 0.07 & 0.12 & 0.10 & 43 \\
19 & 13.17 & 0.52 & 11.32 & 11.32 & 0.00 & 0.00 & 0.00 & 0.00 & 43 \\
20 & 13.16 & 0.51 & 11.70 & 10.63 & 0.22 & 0.04 & 0.21 & 0.22 & 33 \\
21 & 13.15 & 0.51 & 11.82 & 10.87 & 0.53 & 0.53 & 0.23 & 0.47 & 18 \\
22 & 13.12 & 0.50 & 11.57 & 11.25 & 0.32 & 0.20 & 0.31 & 0.26 & 42 \\
\hline
\end{tabular}
}
\tablefoot{From left to right, the columns list the cluster identifier within the corresponding simulation ($gn$), its virial mass ($M_{\rm 200}$) and virial radius ($R_{\rm 200}$), the stellar masses of the central and second most massive galaxies ($M_{\star,\rm 0}$ and $M_{\star,\rm sec}$, respectively), the 3D distance between these galaxies ($d$), their projected distances in the three coordinate planes ($d_{\rm xy}$, $d_{\rm xz}$, and $d_{\rm yz}$), and the number of member galaxies in the cluster ($N$). All of these Fornax-like clusters were selected at redshift $z=0$ from the simulations.}
\label{tab:clusters_dist_massive_galaxies}
\end{table*}
\endgroup

\section{Some details about the \skirt\ implementation adopted in this work}
\label{appendix:skirt}

The wavelength grid used by \skirt\ can be configured manually, or can be read from a file. The first option is recommended when simulating observations at a few discrete wavelengths without taking into account transmission curves, or when working with a grid of many discrete wavelengths with a particular distribution (a logarithmic wavelength grid, for example). In this work, we decided to use the 12 S-PLUS bands with their respective transmission curves (see Sect.~\ref{sec:Fornax-SPLUS}), so the \skirt\ runs were configured to read these curves from data files, with a different file for each survey band. These files must have two columns, one for the wavelength $\lambda$ and another for the transmission corresponding to each $\lambda$, and can be found and downloaded directly from the S-PLUS collaboration website{\footnote{\url{https://www.splus.iag.usp.br/instrumentation/}}}. Strictly speaking, the instrument system simulated by \skirt\ uses a ``default'' wavelength grid, and then each simulated instrument can be configured with its own grid. For simplicity, in this work all wavelength grids were configured in the same way, as established above.

 Regarding the number of photon packets used (photon packets being the fundamental entities for \skirt\ simulations, and which precisely represent the radiation emitted by the sources), in this work it was decided to use $4\times 10^6$ photon packets for each wavelength of the grid, for each of the stellar sources. As shown by \citet{Trayford2017}, this number is adequate for the analysis carried out in this work, being a reasonable balance between the accuracy of the \skirt\ simulations and the computational cost. Since the implementation used here adopts a wavelength grid defined by the 12 S-PLUS photometric bands along with their transmission curves, \skirt\ launches $4\times 10^6$ photon packets for each of the wavelengths that make up each filter, eliminating (if necessary) repeated wavelengths.

The data required by \skirt\ to simulate the stellar sources and their corresponding modeling can be extracted directly from the databases of the corresponding simulations, or can be calculated relatively easily based on them. The quantities needed to trace the emission of the stellar populations used in this work are described below. 

\subsubsection*{Young stellar populations}

To apply the {\sc{mappings-iii}} models, the code requires, for each stellar particle, its coordinates and smoothing length, star formation rate, metallicity, initial mass, and initial density of the gas that gave birth to them. 

\begin{itemize}
    \item {\bf{Star formation rate:}} The {\sc{mappings-iii}} spectra assume a constant star formation rate $\dot{M}_\star$ over the last $10~\rm{Myr}$ for star-forming clouds. Considering a young star particle located within a star-forming cloud and having an initial mass $m_0$ (being such data available in the simulations particle catalogs), the associated star formation rate was defined as $\dot{M}_\star=10^{-7}~m_0~{\rm{yr}}^{-1}$. This value is suitable both to modeling the star formation rate of the gas in the cloud, and to preserving a reasonable value for the stellar mass of the particle (e.g. \citealp{Groves2008}; \citealp{Schaye2015}; \citealp{RodriguezGomez2019}).

    \item {\bf{Metallicity:}} In both \eagle\ and \tng, each star particle has its associated metallicity (defined as $M_{\rm Z}/M_{\rm{total}}$, where $M_{\rm Z}$ and $M_{\rm{total}}$ are, respectively, the mass in elements heavier than He, and the total mass of the particle), which can be extracted directly from the simulation databases.

    \item {\bf{Pressure:}} The {\sc{mappings-iii}} models require the gas pressure in the formation cloud in which the star particles are located ($P_0$). Following \citet{Trayford2017} for \eagle\ galaxies, this was calculated from the density of the gas particle from which the star particle formed, just before it transformed into a star particle (data available in the \eagle\ particle catalogs, in CGS system units), using a polytropic equation of state that limits the star-forming gas pressure (\citealp{DallaVecchia2012}). This leads to $P_0~[\rm{Pa}]=1.38\times10^{-14}\ (1.66\times10^{18}\ \rho_{\rm birth} )^{4/3}$, where $\rho_{\rm birth}$ is the gas density at the time the star particle was born, while the first and second factors involve, respectively, the Boltzmann constant $k_{\rm B}$ and conversions of units (see \citealp{DallaVecchia2012} for details). In contrast, and following the guidelines of \citet{RodriguezGomez2019}, for \tng\ galaxies it was used that $\log[(P_0/k_{\rm B})/\rm{cm}^{-3}~\rm{K}]=5$, or, equivalently, $P_0~[\rm{Pa}]=1.38\times10^{-12}$, being this a typical fixed value for star-forming regions (\citealp{Groves2008}).

    \item {\bf{Compactness:}} The compactness parameter $C$ is a measure of the density of a H{\sc{ii}} region. For the \eagle\ simulations, this parameter was calculated from (\citealp{Groves2008}):
    \begin{equation}
        \log(C)=\dfrac{3}{5}\log\left(\dfrac{M_{\rm cl}}{\rm{M}_\odot}\right)+\dfrac{2}{5}\log\left(\dfrac{P_0/k_{\rm B}}{\rm{cm}^{-3}~\rm{K}}\right) \ ,
    \end{equation}
    \noindent
    being $M_{\rm{cl}} \equiv m_0$ the stellar mass of the cluster in which the cloud forms, and $P_0$ its gas pressure as calculated before. In contrast, for \tng\ galaxies, it was adopted the fixed and typical value $\log(C)=5$ (\citealp{Groves2008}). 

    \item {\bf{Cloud covering factor:}}
    photodissociation regions (PDRs) linked to H{\sc{ii}} regions are far below the resolution of hydrodynamical simulations used in this work, dispersing over time as O and A stars end their lives. The star-forming cloud coverage factor, $f_{\rm PDR}$, gives an estimate of how far the photodissociation region extends relative to the total cloud size. For \eagle, $f_{\rm PDR}=0.1$ was used in this work (\citealp{Trayford2017}), while for \tng, $f_{\rm PDR}=0.2$ was adopted (\citealp{Groves2008}; \citealp{RodriguezGomez2019}).

\end{itemize}

\subsubsection*{Older stellar populations}

For the {\sc{galaxev}} models, \skirt\ requires the coordinates, smoothing length, initial mass, metallicity, and age of the stellar particles as input data. All of these quantities can be directly extracted from particle catalogs of the simulations, for both \eagle\ and \tng.

\end{appendix}

\end{document}